\documentstyle[12pt]{article}

\def\hybrid{\topmargin 0pt      \oddsidemargin 0pt
        \headheight 0pt \headsep 0pt
        \voffset=-0.5cm
        \textwidth 6.5in        % US paper
        \textheight 9in         % US paper
        \marginparwidth 0.0in
        \parskip 5pt plus 1pt   \jot = 1.5ex}
\catcode`\@=11
\def\marginnote#1{}

\newcount\hour
\newcount\minute
\newtoks\amorpm
\hour=\time\divide\hour by60
\minute=\time{\multiply\hour by60 \global\advance\minute by-\hour}
\edef\standardtime{{\ifnum\hour<12 \global\amorpm={am}%
        \else\global\amorpm={pm}\advance\hour by-12 \fi
        \ifnum\hour=0 \hour=12 \fi
        \number\hour:\ifnum\minute<10 0\fi\number\minute\the\amorpm}}
\edef\militarytime{\number\hour:\ifnum\minute<10 0\fi\number\minute}

\def\draftlabel#1{{\@bsphack\if@filesw {\let\thepage\relax
   \xdef\@gtempa{\write\@auxout{\string
      \newlabel{#1}{{\@currentlabel}{\thepage}}}}}\@gtempa
   \if@nobreak \ifvmode\nobreak\fi\fi\fi\@esphack}
        \gdef\@eqnlabel{#1}}
\def\@eqnlabel{}
\def\@vacuum{}
\def\draftmarginnote#1{\marginpar{\raggedright\scriptsize\tt#1}}

\def\draftlabel#1{{\@bsphack\if@filesw {\let\thepage\relax
   \xdef\@gtempa{\write\@auxout{\string
      \newlabel{#1}{{\@currentlabel}{\thepage}}}}}\@gtempa
   \if@nobreak \ifvmode\nobreak\fi\fi\fi\@esphack}
        \gdef\@eqnlabel{#1}}
\def\@eqnlabel{}
\def\@vacuum{}
\def\draftmarginnote#1{\marginpar{\raggedright\scriptsize\tt#1}}

\def\draft{\oddsidemargin -.5truein
        \def\@oddfoot{\sl preliminary draft \hfil
        \rm\thepage\hfil\sl\today\quad\militarytime}
        \let\@evenfoot\@oddfoot \overfullrule 3pt
        \let\label=\draftlabel
        \let\marginnote=\draftmarginnote
   \def\@eqnnum{(\theequation)\rlap{\kern\marginparsep\tt\@eqnlabel}%
\global\let\@eqnlabel\@vacuum}  }

\def\numberbysection{\@addtoreset{equation}{section}
        \def\theequation{\thesection.\arabic{equation}}}

\def\underline#1{\relax\ifmmode\@@underline#1\else
        $\@@underline{\hbox{#1}}$\relax\fi}

\def\titlepage{\@restonecolfalse\if@twocolumn\@restonecoltrue\onecolumn
     \else \newpage \fi \thispagestyle{empty}\c@page\z@
        \def\thefootnote{\fnsymbol{footnote}} }

\def\endtitlepage{\if@restonecol\twocolumn \else  \fi
        \def\thefootnote{\arabic{footnote}}
        \setcounter{footnote}{0}}  %\c@footnote\z@ }
\relax

\numberbysection
\hybrid

\def\beq{\begin{equation}}
\def\eeq{\end{equation}}
\def\p{\partial}
\def\G{\Gamma}
\def\g{\gamma}
\def\s{\sigma}

\def\L{{\cal L}}

\def\C{{\cal C}}

\def\a{\alpha}
\def\b{\beta}
\def\e{\varepsilon}
\def\l{\lambda}

\def\A{{\cal A}}

\def\AD{\A^D_{\g,\a}}
\def\V{{\cal V}}

\def\F{{\cal F}}

\def\L{{\cal L}}

\def\M{{\cal M}}
\def\N{{\cal N}}
\def\O{{\cal O}}
\def\P{{\cal P}}
\def\SP{{\cal S}}

\def\dim{{\rm dim}}
\def\res{{\rm res}}
\def\F{{\cal F}}

\def\wt{\widetilde}
\def\wh{\widehat}

\def \matrix #1 {\left(\begin{array}{cc} #1 \end{array}\right)}
\newtheorem{th}{Theorem}[section]

\newtheorem{cor}{Corollary}[section]
\newtheorem{lem}{Lemma}[section]

\begin{document}

\input epsf

\begin{titlepage}
\title{Isomonodromy equations on algebraic curves, canonical transformations
and Whitham equations}

\author{I.Krichever \thanks{Columbia University, New York, USA and
Landau Institute for Theoretical Physics and ITEP, Moscow, Russia; e-mail:
krichev@math.columbia.edu. Research is supported in part by National Science
Foundation under the grant DMS-01-04621 and by CRDF Award RP1-2102}}

\date{December 4, 2001}

\maketitle
The Hamiltonian theory of isomonodromy equations for meromorphic
connections with irregular singularities on algebraic curves is
constructed. An explicit formula for the symplectic structure on
the space of monodromy and Stokes matrices is obtained. The Whitham
equations for the isomonodromy equations are derived. It is shown
that they provide a flat connection on the space of the spectral curves
of the Hitchin systems.

\begin{abstract}

\end{abstract}
\vfill

\end{titlepage}
\newpage

\section{Introduction}

The goal of this paper is multi-fold. Our first objective is to construct
isomonodromy equations for meromorphic connections with irregular and
regular singularities on algebraic curves. The isomonodromy equations for
linear systems with irregular singularities on rational curve generalizing
Schlesinger's equations \cite{schles} were introduced by Jimbo,
Miwa and Ueno \cite{jimbo}. A particular case of these equations
was considered earlier by Flashcka and Newell \cite{flashka} in
connection with a theory of self-similar solutions of the mKdV
equation. Fuchsian systems on higher genus Riemann surfaces
were considered in \cite{iwasaki}. The case of linear systems with one
irregular singularity on an elliptic curve in \cite{okomoto}.
The recent burst of interest to the isomonodromy
equations for linear systems with regular singularities
on higher genus Riemann surfaces is due to their connections with
the classical limit of Knizhnik-Zamolodchikov-Bernard equations for
correlation functions of the Wess-Zumino-Witten-Novikov theory.
For the case of rational and elliptic curves these connections
were revealed in \cite{resheto,harnard,korotkin}. General case was
considered in \cite{levin}, where more complete list of references
can be found.

The conventional modern approach to a theory of the isomonodromy
equations is based on their representation in a form of compatible
non-autonomous Hamiltonian systems that can be identified as the
Hamiltonian reduction of some free Hamiltonian theory. This approach
presents an almost exhaustive geometric description of the system but it
requires solving the corresponding moment map equations in order to get
an explicit form of the equations or their Lax representation. The moment map
equations are differential equations on the algebraic curve. They
have been solved explicitly only in very few cases \cite{levin}.

As in \cite{schles,jimbo}, the starting point of our
approach is the Lax representation of the isomonodromy equations.
In the next section the space of meromorphic connections on stable,
rank $r$, and degree $rg$ holomorphic vector bundles on an algebraic curve
$\G$ with the poles divisor $D=\sum_m (h_m+1)P_m$
is identified with orbits $\A^D/SL_r$ of the adjoint action of
$SL_r$ on a certain subspace $\A^D$ of meromorphic matrix-valued
differentials on $\G$. A characteristic property of $\wt L\in \A^D$
is that all its additional singularities at points $\g_s\notin D$
have the form $d\Phi \Phi^{-1}$, where $\Phi$ is holomorphic.
We show that an open set of $\A^D$, corresponding to the case when
all the additional poles of $\wt L$ are simple, is parametrized by the data
\begin{eqnarray}
\wt L_m, \ (\g_s,\kappa_s),\ L_{s0}=\b_s\a_s^T, \label{01}\\
\sum_{s=0}^{rg}L_{s0}+\sum_m \res_{P_m} \wt L_m=0,\nonumber
\end{eqnarray}
where $\wt L_m$ is the singular part of $\wt L$ at $P_m$,
$(\g_s,\kappa_s)$ is a point of the bundle of scalar affine connections
on $\G$, and  $L_{s0}$ is a {\it rank} 1 matrix such that
${\rm Tr} \ L_{s0}=1$.
We identify matrices $L_{s0}$ with pairs of $r$-dimensional vectors
$\a_s=(\a_s^i), \ \b_s=(\b_s^i)$, considered modulo transformation
$\a_s\to \lambda_s\a_s,\ \b_s\to \lambda_s^{-1}\b_s$, and such that
$(\a_s^T\b_s)=1$.

>From the definition of $\wt L\in \A^D$ it follows that the equation
\beq\label{sys0}
d\Psi=\wt L\Psi
\eeq
has a multi-valued holomorphic solution on $\G\backslash D$. Let us fix
a point $Q\in \G$. Then, the analytical continuation of $\Psi$, normalized by
the condition $\Psi(Q)=1$, defines a representation of the
fundamental group $\pi_1(\G\backslash D;Q)\longmapsto GL_r$. The Stokes
matrices and
the so-called exponents at the irregular singularities $P_m,\ h_m>0$ can
be defined purely locally, as in the case of genus $g=0$, if a local
coordinate in the neighborhood of $P_m$ is fixed.

The Stokes data and the exponents at $P_m$ depend
only on the $h_m$-jet of the local coordinate, and therefore, we identify the
space of the isomonodromy deformations of the linear system
(\ref{sys0}) with the moduli space $\M_{g,1}(h)$ of smooth genus $g$
algebraic curves with a puncture $Q$, and with fixed $h_m$-jets of local
coordinate at punctures $P_m$. Here and below the {\it isomonodromy
deformations are the ones which preserve the monodromy representation, the
Stokes matrices, and the exponents}. For brevity, all these data
we simply call monodromy data.

It is necessary to emphasize that for $g=0$ our definition of the
deformation space is equivalent to the traditional one. According to
\cite{jimbo,flashka}, the isomonodromy deformations of $\wt L$ are
parameterized by the positions of poles and the exponents at the irregular
singularities. In this setting the local coordinates at the poles are
always fixed, and are defined by the global coordinate on a complex plane.
It is easy to show that the deformations of the exponents that correspond to
gauge invariant equations for $\wt L$ can be identified with deformations
corresponding to a change of the local coordinate.

A change of normalization point $Q$, and the gauge transformation of
$\wt L'=g\wt Lg^{-1},\ g\in GL_r$ correspond to conjugation of the
monodromy data by a constant matrix. Hence, the space of the
isomonodromy deformations of the meromorphic connections $\A^D/SL_r$ is the
moduli space $\M_g(h)$ of curves with $h_m$-jets of local coordinates at
$P_m$. For the connections with regular singularities $h_m\equiv 0$ at $N$
points the deformation space is just $\M_{g,N}$.
The space $\A(h)$ of the all admissible meromorphic differentials
with fixed multiplicities $(h_m+1)$ at the punctures, and the
corresponding factor-space of meromorphic connections we consider as
bundles
\beq\label{00}
\A(h)\longmapsto \M_{g,1}(h),\ \ \A(h)/SL_r\longmapsto \M_g(h).
\eeq
In section 4 the Lax representation for a full hierarchy of the
isomonodromy equations is derived. We show that the Lax equations
are equivalent to a system of well-defined compatible evolution
equations on the space of dynamical variables, which are the parameters
(\ref{01}). Naively, the Lax
representation
\beq\label{lax0}
\p_{T_a}\wt L=[M_a,\wt L]-dM_a,
\eeq
of the isomonodromy equations is just the coordinate-dependent way of
saying that $M_a$ are the coefficients of a flat connection on the space of linear systems (\ref{sys0}) defined by the monodromy data.
In order to make sense out of (\ref{lax0}), it is necessary first to express
$M_a$ as a function of $\wt L$, and to show then, that the Lax equation
is equivalent to a well-defined system of differential equations for $\wt L$.
{\it A'priory} the last statement is not obvious, because (\ref{lax0}) has
to be fulfilled identically on $\G$, and the space of $\wt L$ is
finite-dimensional. For example, for $g>0$ it is impossible to define
the isomonodromy deformations for matrix-valued differentials with
poles only at $D$.  The presence of extra poles $\g_s$, which become dynamical
variables, is a key element, which allows us to overcome that difficulty
in defining the isomonodromy equations on higher genus algebraic curves.
The very same idea was used in our earlier work \cite{kr0},
where an explicit parameterization of the Hitchin systems \cite{hitchin}
was obtained, and where infinite-dimensional field generalizations
of the Hitchin systems were proposed.

In Section 5 we show that the approach to the Hamiltonian theory of
soliton equations proposed in \cite{kp1,kp2,kr4} is also applicable
to the case of isomonodromy deformations. The key element of this approach
is a definition of the universal two-form which is expressed in terms of the
Lax operator and its eigenvectors. The proof that the contraction of
this form by the vector-field defined by a Lax equation is an exact one-form
is very general and does not rely on any specific form of the Lax operator.
It provides a direct way to show that the Lax equations are Hamiltonian
on suitable subspaces, and at the same time allows to identify the
corresponding Hamiltonians.

It turns out that the universal two-form on a space of meromorphic
connections is defined identically to the case of isospectral equations
if we replace eigenvectors by a solution of equation (\ref{sys0}).
More precisely, let $\P_0$ be a subspace of $\A(h)$ with fixed exponents at
the punctures, and let $\psi_m$ be the formal local solutions of (\ref{sys0})
at $P_m$ (see (\ref{f3}) below). Then the formula
\beq\label{f1}
\omega=-{1\over 2}\sum_{s=1}^{rg}\res_{\g_s}
{\rm Tr} \left(\psi^{-1} \delta \wt L\wedge \delta \psi\right)-{1\over 2}
\sum_{P_m}\res_{P_m} {\rm Tr} \left(\psi_m^{-1} \delta \wt L\wedge \delta
\psi_m\right)
\eeq
defines a closed, non-degenerate form on the factor space $\P=\P_0/SL_r$.
The Lax equations restricted to $\P_0$ descends to a system of commuting
flows which are Hamiltonian with respect to the symplectic
structure defined by $\omega$.

We show that $\omega$ in terms of the parameters (\ref{01})
can be written as
\beq\label{03}
\omega=\sum_{s=1}^{rg}\left(\delta \kappa_s\wedge\delta z_s+\sum_{i=1}^r
\delta \b_s^i \wedge\delta \a_s^i\right)+\sum_m \omega_m\ ,
\eeq
where $\omega_m$ is the canonical symplectic structure on an orbit $\wt \O_m$
of the adjoint action of the group of invertible formal holomorphic matrix
functions on the space of singular parts of meromorphic matrix differentials
in a formal disc with the pole of order $h_m$. (A set of orbits $\wt \O_m$
corresponds to the set of fixed exponents.)

A remarkable property of the symplectic structure for isospectral equations,
defined in terms of the Lax operator is that it provides, under quite
general circumstances, a straightforward way of construction of
action-angle type variables (see examples in \cite{kp1}-\cite{kp4}).
In section 6 we show that in the case of the isomonodromy
equations almost the same arguments lead to an expression of the
symplectic form $\omega$ in terms of the monodromy data.

For example, the monodromy data corresponding to a meromorphic connection
on an elliptic curve are just a pair of matrices $A$ and $B$, considered
modulo common conjugation. The monodromy matrix around the puncture is
equal to
\beq
J=B^{-1}A^{-1}BA\  .
\eeq
Symplectic leaves $\P$ are defined by a choice of the orbit for $J$.
Therefore, they can be seen as level sets of the invariants
${\rm Tr}\  J^k$. We show that the symplectic form on $\P$, defined by
$\omega$ is equal to the restriction onto $\P$ of the two-form
\beq\label{x}
\chi(A,B)={\rm Tr} \left[B^{-1}\delta B\wedge \delta A A^{-1}-
A^{-1}\delta A \wedge \delta B B^{-1}+\delta J J^{-1}\wedge
B^{-1}A^{-1}\delta (AB) \right] .
\eeq
The expression for $\omega$ on symplectic leaves of the space of conjugacy
classes of representation of the fundamental group of genus $g$
Riemann surface with one puncture is given by the formula (\ref{ct7}).
In a different form this result was obtained in \cite{goldman}.
An $r$-matrix representation of the Poisson structure on the space of flat
connections on Riemann surfaces with boundaries was found in \cite{fock}.

To the best of the author's  knowledge, the general closed expression
for the symplectic structure on orbits of the adjoint action of $SL_r$
on the space of monodromy matrices $A_i,B_i$ and Stokes matrices,
given by the Theorem 6.1 is new. Even in the genus 0 case, the Poisson structure
on the space of Stokes matrices corresponding to meromorphic connections
with one irregular singularity of order 2 and one regular singularity was
found only recently \cite{boalch}. The Poisson structure was identified with
that of the Poisson-Lie group $G^*$ dual to $G=GL_r$.
In \cite{boalch1} this result
was generalized for $G$-valued Stokes matrices for arbitrary simple Lie
group, and very interesting connections with a theory of Weyl quantum
groups was found.  The Poisson structure on the space of Stokes matrices
corresponding to skew-symmetric meromorphic connection with one regular and
one irregular order 2 singularity was found earlier in \cite{ugaglia}.
The Poisson structure for $(2\times 2)$ Stokes matrices corresponding
to meromorphic connections with one irregular singularity of order 4 was
obtained in \cite{flashka}.

To some extend our main result of Section 6 is preliminary. The general
expression for $\omega$ in terms of the monodromy data came out of the blue,
as a result of straightforward computations. It seems urgent to find its
interpretation in terms of the Poisson-Lie group theory.

The last goal of this paper is to establish connections between solutions of
the isomonodromy equations on algebraic curves and solutions of the Hitchin
systems. It is well-known that solutions of the Schlesinger equations after
proper rescaling can be treated as  ``modulation'' of solutions of the Garnier
system \cite{garnier}. An attempt to revisit this connection in light of the
Whitham theory \cite{ffl,kr1,kr2,kr5} was made in \cite{takasaki}, but the
heuristic arguments used in \cite{takasaki} do not allow to derive the
modulation equations in a closed form.

The problem which we address in Section 7 is as follows.
The space of meromorphic $\e$-connection with fixed
multiplicities $h=\{h_m\}$ of poles is the space of orbits of the adjoint
action of $SL_r$ on the space $\A_{\e}(h)$ of meromorphic differentials
$\wt L_{\e}$ such that $\e^{-1}L_{\e}\in \A(h)$. They are parametrized
by the data (\ref{01}) such that ${\rm Tr}\  L_{s0}=\e$.
The family of meromorphic $\e$-connections defined for $\e\neq 0$
extends to a smooth family over the whole disc. The central fiber over $\e=0$
parametrizes the space $\L$ of Lax matrices on algebraic curves introduced
in \cite{kr0}. The orbits of the adjoint action of $SL_r$
on a subspace of $\L$, corresponding to a {\it fixed} algebraic curve $\G$,
and fixed singular parts of the eigenvalues of $\wt L_m$ can be identified
with the phase space of the generalized Hitchin system.

In order to get a smooth at $\e=0$ family of the isomonodromy
equations for $\e$-connections, it is necessary to rescale the coordinates
$T_a$ on $\M_{g,1}(h)$. More precisely, if we define the coordinates
$t_a=\e^{-1} T_a$, then the deformations of $\wt L_{\e}$ that preserve
the monodromy data associated with a solution of the equation
\beq\label{0sys}
\e d\psi=\wt L_{\e}\psi
\eeq
are described by the equations
\beq\label{0lax}
\p_{t_a}\wt L_{\e}-\e dM_a+[\wt L_{\e},M_a]=0.
\eeq
The equations (\ref{0lax}) are Hamiltonian and the corresponding
Hamiltonians do converge to certain {\it quadratic} Hamiltonians of
the Hitchin system, as $\e\to 0$. Therefore, locally solutions of
(\ref{0lax}) converge to the solutions of the Hitchin system.
At the same time a global behaviour of solutions of the isomonodromy and
isospectral flows is quit different. The monodromy data which are preserved
by (\ref{0lax}) vanish in the limit $\e\to 0$. The space of integrals of the
Hitchin system can be regarded as a space $\SP$ of the so-called spectral
curves. It is of dimension which is only half of the dimension
the monodromy data. For $\wt L_0\in \L=\A_0(h)$ the time-independent
spectral curve is defined by the characteristic equation
\beq\label{car}
\det \ (\wt k-\wt L)=0 .
\eeq
The spectral curve $\wh \G$ is $r$-fold branch cover of the initial
algebraic curve $\G$. The equations of motion for the Hitchin system
are linearized on the Jacobian of $\wh \G$.

In Section 7 we apply ideas of the multi-scale perturbation theory
to the construction of asymptotic solutions of the isomonodromy equations
using solutions of the Hitchin system. In this approach the
leading term of the approximation describes the motion  which is to first
order the original fast motion on the Jacobian, combined with a slow drift on
the moduli space of the spectral curves. We obtain an explicit form
of the Whitham equations describing that slow drift. They imply that
{\it the real part of the periods  of the differential} $\wt k$ on $\wh \G$
{\it are preserved along the slow drift}. We would like to emphasize
that the correspondence
\beq\label{09}
\wh \G\in \SP\longmapsto {\rm Re }\oint_c \wt k , \ \ \ c\in H_1(\wt \G)
\eeq
define a flat {\it real} connection on the moduli space of the spectral
curves, considered as a bundle over $\M_g(h)$.
To some extend our result provides evidence that this connection is
a residual of the flat connection on $\A_{\e}(h)$ defined by the monodromy
data in the limit $\e\to 0$. It would be quite interesting
to find a more geometrical interpretation of that residual correspondence.

\section{Meromorphic connections}

Let $V$ be a stable, rank $r$, and degree $rg$ holomorphic vector bundle
on a smooth genus $g$ algebraic curve $\G$. Then the dimension of the space
of its holomorphic sections is $r=\dim \ H^0(\G,V)$. Let $\s_1,\ldots,\s_r$
be a basis of this space. The vectors $\s_i(\g)$ are linear independent
at the fiber of $V$ over a generic point $\g\in \G$, and are linearly dependent
\beq\label{a}
\sum_{i=1}^{r} \a_{s}^i \s_i(\g_s)=0
\eeq
at zeros $\g_s$ of the corresponding section of the determinant bundle
associated to $V$. For a generic $V$ these zeros are simple, i.e.
the number of distinct points $\g_s$ is equal to $rg=\deg \ V$, and
the vectors $\a_s=(\a_s^i)$ of linear dependence (\ref{a})
are uniquely defined up to a multiplication.
A change of the basis $\s_i$ corresponds to a linear transformation of
$\a_s'=g^T\a_s$. Hence, an open set $\M\subset \wh \M$ of the moduli space
of vector bundles is parameterized by points of the factor-space
\beq\label{mod}
\M=\M_0/SL_r,\ \ \M_0\subset S^{rg}\left(\G\times CP^{r-1}\right) \ ,
\eeq
where $SL_r$ acts diagonally on the symmetric power of $CP^{r-1}$.
In \cite{kn1,kr3} the parameters $(\g_s,\a_s)$ were called Tyurin parameters.

Let  $(\g,\a)=\{\g_s,\a_s\}$ be a point of the symmetric product
$X=S^{rg}\left(\G\times CP^{r-1}\right)$. Throughout the paper it is
assumed that the points $\g_s\in \G$ are distinct, $\g_s\neq \g_k$.
The vector bundle $V_{\g,\a}$ corresponding  to $(\g,\a)$ under the
inverse to the Tyurin map is described in terms of Hecke modification of the
trivial bundle. In this description the space of local sections of the vector
bundle $V_{\g,\a}$ is identified with the space $\F_s$ of meromorphic
(row)vector-functions in the neighborhood of $\g_s$ that have simple pole
at $\g_s$ of the form
\beq\label{F}
f^T(z)={\l_s\a_s^T\over z-z(\g_s)}+O(1),\ \ \l_s\in C.
\eeq
Our next goal is to describe in similar terms the space of meromorphic
connections on $V_{\g,\a}$. Let $D=\sum_m (h_m+1)P_m$ be an effective divisor
on $\G$ that does not intersect with $\g$. Then we define
the space $\AD$ of meromorphic matrix valued differentials
$\wt L=L(z)dz$ on $\G$ such that:

$1^0.$ $\wt L$ is holomorphic except at the points $\g_s$, where it has at most
simple poles, and at the  points $P_m$ of $D$, where it has poles of degree
not greater than $(h_m+1)$;

$2^0.$ the singular term of the expansion
\beq\label{Ls}
\wt L=\left({L_{s0}\over z-z_s}+L_{s1}+L_{s2}(z-z_s)+O((z-z_s)^2)\right)dz,
\ \ z_s=z(\g_s),
\eeq
is a rank 1  matrix of the form
\beq \label{Ls0}
L_{s0}=\b_s\a_s^T\ \longleftrightarrow L_{s0}^{ij}=\\b_s^i\a_s^j,
\eeq
where $\b_s$ is a vector. Trace of the residue of $\wt L$ at $\g_s$ equals
$1$:
\beq\label{1}
\res_{g_s} {\rm Tr} \ \wt L=1 \ \ \longmapsto
\a_s^T\b_s={\rm tr}\ L_{s0}=1;
\eeq

$3^0.$ $\a_s^T$ is a left eigenvector of the matrix $L_{s1}$, i.e.
\beq\label{Ls1}
\a_s^TL_{s1}=\a_s^T\kappa_s,
\eeq
where $\kappa_s$ is a scalar.

Note that the condition $(3^0)$ is well-defined, although the expansion
(\ref{Ls}) by itself does depend  on the choice of a local coordinate $z$
in the neighborhood of $\g_s$. Under a change of local coordinate
$w=w(z)$ the eigenvalue $\kappa_s$ in (\ref{Ls1}) gets transformed to
$\kappa_s'$, where
\beq\label{affine}
\kappa_s=\kappa_s' w'(z_s)-{w''(z_s)\over 2w'(z_s)}\ .
\eeq
Therefore, the pair $(\g_s,\kappa_s)$ is a well-defined point of
a total space of the bundle $C^{aff}(\G)$ of scalar affine connections on $\G$.

Sum of all the residues of a meromorphic differential equals zero. Therefore,
\beq\label{2}
\sum_{P_m\in D}\res_{P_m} {\rm Tr} \ \wt L=-rg.
\eeq
Hence, in what follows we always assume that $\deg D=N>0.$ The Riemann-Roch
theorem implies that for a generic degree $N$ divisor $D$ and a generic set of
Tyurin parameters $(\g,\a)$ the space $\AD$ is of dimension
\beq\label{dim}
\dim\  \AD=r^2(N+rg+g-1)-r^2g(r-1)-rg-rg(r-1)=r^2(N+g-1).
\eeq
The first term is the dimension of the space of meromorphic differentials on
$\G$ with the pole divisor $D+\g$. The consecutive terms count the numbers
of the constraints (\ref{Ls}-\ref{Ls1}).
A key characterization of these constraints is as follows.
\begin{lem}
A meromorphic matrix-function $L$ in the neighborhood $U$ of $\g_s$
with a pole at $\g_s$ satisfies the constraints (\ref{Ls}-\ref{Ls1})
if and only if it has the form
\beq\label{lgauge}
\wt L=d\Phi_s(z)\Phi_s^{-1}(z)+\Phi_s(z)\wt L_s(z)\Phi_s^{-1}(z),
\eeq
where $\wt L_s$ and $\Phi_s$ are holomorphic in $U$, and
$\det \Phi_s$ has at most simple zero at $\g_s$.
\end{lem}
The proof is almost identical to that of the Lemma 2.1 in \cite{kr0}.

The constraints (\ref{Ls}-\ref{Ls1}) imply that the space $\F_s$ is
invariant under the adjoint action of the operator $(\p_z-L)$, i.e.
\beq\label{action}
f^T\in \F_s\longmapsto f^T(\p_z-L)=-\left(\p_zf^T+f^TL\right)\in \F_s.
\eeq
Therefore, for generic set of the Tyurin parameters $(\g,\a)$
the factor-space $\AD/SL_r$ corresponding to the gauge transformations
\beq\label{gauge}
\wt L\to g\wt Lg^{-1},\ \ g\in SL_r,
\eeq
can be identified with the space of meromorphic
connections on $V_{\g,\a}$ that have poles at
$P_m$ of degree not greater that $h_m+1$.

The explicit parameterization of an open set of the phase space of
the Hitchin systems proposed in \cite{kr0} can be easily extended to the case
under consideration. Consider first an open set of the Tyurin parameters
such that the dimension of the space $\F_{\g,\a}$ of meromorphic
(row)vector-functions on $\G$ with simple poles at $\g_s$ of the
form (\ref{F}) equals $r$. Then, as shown in \cite{kr0} the
matrix $\a_s^i$ is of rank $r$. We call $(\g,\a)$ a non-special set of the
Tyurin parameters if additionally they satisfy the constraint:
there is a subset of $(r+1)$ indices $s_1,\ldots,s_{r+1}$ such that
all minors of $(r+1)\times r$ matrix $\a_{s_j}^i$ are non-degenerate.
The action of the gauge group on the space of non-special sets of the Tyurin
parameters $\M_0$ is free. We also assume that the corresponding
points $\g_s$ do not coincide with the points $P_m$.

By definition, the singular part $\wt L_m$ of a meromorphic differential
$\wt L$ is an equivalence class of meromorphic differentilas in the
neighborhood of $P_m$ considered modulo holomoprhic differentials.
\begin{lem} Let $\A^D$ be an affine bundle over $\M_0$ with fibers
$\AD$. Then the map
\beq\label{100}
\wt L\in \A^{D}\longmapsto \{\a_s,\b_s,\g_s,\kappa_s,\wt L_m\}\, ,
\eeq
is a bijective correspondence between points of the bundle
$\A^D$ over $\M_0$ and sets of the data (\ref{100}) subject to the
constraints
$(\a_s^T\b_s)=1$, and
\beq\label{B2}
\sum_{s=1}^{rg} \b_s\a_s^T+\sum_{P_m\in D'}\res_{P_m} \ \wt L_m=0,
\eeq
modulo gauge transformations
\beq\label{101}
\a_s\to \l_s \a_s,\ \ \b_s\to \l_s^{-1}\b_s\, .
\eeq
\end{lem}
Recall that we consider the pairs $(\g_s,\kappa_s)$ as points of the bundle
$C^{aff}(\G)$.

\medskip
\noindent{\bf Example.} Let $\G$ be a hyperelliptic curve
defined by the equation
\beq\label{hyp}
y^2=R(x)=x^{2g+1}+\sum_{i=0}^{2g}u_ix^i
\eeq
Parametrization of connections on $\G$ with simple pole at the infinity
is almost identical to parameterization of the Hitchin systems on $\G$
proposed in \cite{kr0}.
A set of points $\g_s$ on $\G$ is a set of pairs $(y_s,x_s)$, such that
\beq\label{hyp1}
y_s^2=R(x_s)\ .
\eeq
A meromorphic differential on $\G$ with residues $(\b_s\a_s^T)$ at $\g_s$
and a simple pole at the infinity has the form
\beq\label{hyp2}
L{dx\over 2y}=\left(\sum_{i=0}^{g-1}L_i x^i+\sum_{s=1}^{rg}(\b_s\a_s^T) \
{y+y_s\over x-x_s}\right)\ {dx\over 2y}\ ,
\eeq
where $L_i$ is a set of arbitrary matrices. The constraints (\ref{Ls1})
are a system of linear equations defining
$L_i$:
\beq\label{hyp3}
\sum_{i=0}^{g}\a_n^TL_i x_k^i+\sum_{s\neq n}(\a_n^T\b_s)\a_s^T \
{y_n+y_s\over x_n-x_s}=\kappa_n \a_n^T,\ \ n=1,\ldots,rg,
\eeq
in terms of  data $\{\g_s,\kappa_s, \a_s,\b_s\}$, where $(\a_s,\b_s)$ are
arbitrary vectors such that $\a_s^T\b_s=1$.

For $g>1$, the correspondence (\ref{100}) descends to a system of local
coordinates on $\A^{D}/SL_r$.
Consider the open set of $\M_0$ such that the vectors
$\a_j,\  j=1,\ldots, r,$ are linearly independent and all the coefficients
of an expansion of $\a_{r+1}$ in this basis do not vanish
\beq\label{pa33}
\a_{r+1}=\sum_{s=1}^r c_j \a_j,\ \ c_j\neq 0.
\eeq
Then for each point of this open set there exists a unique matrix
$W\in GL_r$, such that $\a_j^TW$ is proportional to the basis vector $e_j$
with the coordinates $e_j^i=\delta_j^i$, and $\a_{r+1}^TW$ is proportional to
the vector $e_0=\sum_j e_j$.
Using the global gauge transformation defined by $W$
\beq\label{pa34}
b_s=W^{-1}\b_s,\ \ a_s=W^T\a_s,
\eeq
and the part of local transformations
\beq \label{local}
a_s\to \lambda_s a_s;\ a_s\to \lambda_s^{-1}b_s,
\eeq
for $s=1,\ldots,r+1,$ we obtain that on the open set of $\M_0$
each equivalence class has representation of the form
$(a_s,b_s)$ such that
\beq\label{AA}
a_i=e_i,\ i=1,\ldots,r;\ a_{r+1}=e_0.
\eeq
This representation is unique up to local transformations (\ref{local})
for $s=r+2,\ldots,rg$.

In the gauge (\ref{AA}) equation (\ref{B2}) can be easily solved for
$b_1,\ldots,b_{r+1}$. Using (\ref{AA}), we get
\beq\label{pa35}
b_j^i+b_{r+1}^i=-\sum_{s=r+2}^{rg}b_s^ia_s^j-\sum_m \res\  \wt L_m^{ij}\ .
\eeq
The condition $a^T_jb_j=1$ for $a_j=e_j$ implies $b_j^j=1$.
Hence,
\beq \label{pa36}
b_{r+1}^i=-1-\sum_{s=r+2}^{rg}b_s^ia_s^i-\sum_m \res\  \wt L_m^{ij}.
\eeq
Note, that the constraint (\ref{2}) implies $a_{r+1}^Tb_{r+1}=1.$

Sets of vectors $a_s, b_s, a_s^Tb_s=1, r+1<s\leq rg$,
modulo the transformations (\ref{local}), and points
$\{\g_s,\kappa_s\}\in S^{rg} \left(C^{aff}(\G)\right)$, and sets $\wt L_m$,
satisfying (\ref{2}), provide a parametrization of an open set of the
bundle $\A^D/SL_r$ over $\M=\M_0/SL_r$. The dimension of this bundle equals
\beq\label{dimA}
\dim\  \A^D/SL_r=r^2(N+2g-2)+1.
\eeq
In the same way, taking various subsets of $(r+1)$ indices we obtain charts
of local coordinates which cover $\A^D/SL_r$.

\section{Monodromy data}

Our next goal is to introduce monodromy data corresponding to $\wt L\in \AD$
along identical lines to their definition in the zero genus case.
>From Lemma 2.1 it follows that the equation
\beq\label{sys}
d\Psi=\wt L\Psi
\eeq
has multi-valued holomorphic solutions on $\G\backslash \{P_m\}, \ P_m\in D$.
Let $Q$ be a point on $\G$. Then the normalization
\beq\label{n0}
\Psi(Q)=1
\eeq
defines $\Psi$ uniquely in the neighborhood of $Q$.
Analytic continuation of $\Psi$ along cycles in $\G\backslash \{P_m\}$
defines the monodromy representation
\beq\label{rep}
\mu:\pi_1(\G\backslash \{P_m\};Q)\longmapsto GL_r .
\eeq
It is well-known, that for connections with simple poles
the correspondence $\wt L\to \mu$ is an injection,
and that the inverse map is defined on an open set of the space of
representations. For connections with poles of higher order
additional so-called Stokes data are needed. Their construction
is local, and we mainly follow here \cite{sib}.
\begin{lem}
Let $L$ be a formal Laurent series
\beq\label{f}
\wt L=\sum_{i=-h}^{\infty} L_{s}w^{s-1}dw
\eeq
such that the leading coefficient has the form
\beq\label{eigen2}
L_{-h}=\Phi K\Phi^{-1}, \ \
K={\rm diag}\ \left(k_{1},\ldots,k_{r}\right),\ \
\left\{\begin{array}{l}k_{i}-k_{j}\neq 0, \ h>0, \\
k_{i}-k_{j}\notin Z,
\ h=0,
\end{array}\right.\ i\neq j.
\eeq
Then the equation (\ref{sys}) has a unique formal solution
\beq\label{f3}
\psi=\Phi\left(1+\sum_{s=1}^{\infty}\xi_sw^s\right)
\exp\left(\ \sum_{i=-h}^{\infty}
K_i\int w^{i-1}dw\right),
\eeq
where $K_i$ are diagonal matrices, $K_{-h}=K$,
and the matrices $\xi_s$ have zero diagonals, $\xi_s^{ii}=0$.
\end{lem}
Substitution of (\ref{f3}) into (\ref{sys}) gives a system of equations, which
for $h>0$ have the form
\beq\label{f4}
K_s+[K,\xi_{s+h}]=R(\xi_1,\ldots,\xi_{s+h-1},K_{-h+1},\ldots,K_{s}),\ s>-h.
\eeq
They recursively determine the off-diagonal part of $\xi_s$ and the
diagonal matrix $K_s$. In the similar way $\psi$ is constructed for $h=0$.

Suppose now that ({\ref{f}) is the Laurent expansion of a meromorphic
differential in a punctured disk $U$ which is holomorphic in
$\wh U=U\backslash 0$.
Let $V$ be sector of $\wh U$ that for any pair $(i,j)$ contains
only one ray, such that
\beq\label{f5}
{\rm Re} \ (k_i-k_j)w^{-h}=0.
\eeq
Then there exists a holomorphic in $V$ solution $\Psi_V$ of (\ref{sys})
such that the formal solution (\ref{f3}) is an asymptotic series for
$\Psi_V$. The asymptotic is uniform in any closed subsector of $V$.

The punctured disk can be covered by a set of sectors $V_1,\ldots, V_{2h+1}$
which satisfy the constraint described above, and such that the sectors
$V_\nu$ and $V_{\nu+1}$ do intersect. On their intersection the
solutions $\Psi_{\nu}=\Psi_{V_{\nu}}$ and $\Psi_{\nu+1}=\Psi_{V_{\nu+1}}$
satisfy the relation
\beq\label{S}
\Psi_{\nu+1}=\Psi_{\nu}S_{\nu}, \ \nu=1,\ldots, 2h.
\eeq
The Stokes' matrices $S_{\nu}$ are constant matrices.
For each $S_\nu$ there exists a unique permutation under which $S_\nu$ gets
transformed to an upper triangular matrix with the diagonal elements equal $1$.

The last property of the Stokes' matrices follows from more precise
statement which we will use in Section 6. Namely, if $w$ tends to $0$
in the intersection of $V_{\nu}$ and $V_{\nu+1}$, then the following
limit exists and equals
\beq\label{S1}
\lim_{w\to 0} \ \exp(Kw^{-h})\ S_{\nu}\exp({-Kw^{-h}})=1.
\eeq
For any pair $(i\neq j)$ the left hand side of (\ref{f5}) has a definite
sign in $V_{\nu}\cap V_{\nu+1}$. Therefore, if this sign is
positive, then (\ref{S1}) implies $S_{\nu}^{ij}=0$.

Let us fix a local coordinate $w_m$ in the neighborhood of $P_m,\ w_m(P_m)=0$,
and paths $c_m$ connecting $Q$ with $P_m$. In the neighborhood of $P_m$ we
also fix a set of sectors $V_{\nu}^{(m)}$ described above, and always assume
that the path $c_m$ in the neighborhood of $P_m$ belongs to the first sector
$V_1^{(m)}$. Then the Laurent expansion of $\wt L\in \A^D$ at $P_m$ in this
coordinate, defines the diagonal matrices $K_i^{(m)}$,  the Stokes' matrices
$S_{\nu}^{(m)}$, and the transition matrice $G_m$,
which connects $\Psi$ and $\Psi_1^{(m)}$
\beq\label{S3}
\Psi=\Psi_{1}^{(m)}G_{m}.
\eeq
In each of the sectors $V_{\nu}^{(m)}$ we have
\beq\label{S4}
\Psi=\Psi_{\nu}^{(m)}g_{\nu}^{(m)},
\eeq
where
\beq\label{S5}
g_{1}^{(m)}=G_{m},\ \
g_{\nu+1}^{(m)}=\left(S_{1}^{(m)} S_2^{(m)}\cdots S^{(m)}_{\nu}\right)^{-1}
G_{m}, \
\nu=1,\ldots,2h_m.
\eeq
The monodromy $\mu_{m}$ around $P_m$ is equal to
\beq\label{f6}
\mu_{m}=\left(g_1^{(m)}\right)^{-1}e^{2\pi iK_0^{(m)}}g_{2h_m+1}^{(m)}=
G_{m}^{-1}e^{2\pi iK_0^{(m)}}
\left(S_{1}^{(m)} S_2^{(m)}\cdots S^{(m)}_{2h_m}\right)^{-1}
G_{m}
\eeq
If we choose a basis of $a_j,b_j$ cycles on $\G$ with the canonical matrix of
intersections, then we denote the monodromy matrices along the cycles by
$A_j,B_j,\ j=1,\ldots,g$.
\begin{lem}
The correspondence
\begin{eqnarray}\label{data}
\wt L\in \A^D\longmapsto
\{K_i^{(m)},\ S_{\nu}^{(m)},\ G_m,\ A_j,\ B_j \},\\
\ -h_m \leq i\leq 0,\ \ \nu=1,\ldots, 2h_m,
\end{eqnarray}
where the transition and the Stokes matrices are considered modulo
transformations
\beq\label{trans}
G_m\longmapsto W_mG_m,\ \ S_{\nu}^{(m)}\longmapsto W_m S_{\nu}^{(m)}
W_m^{-1},\ \ W_m={\rm diag}\ (W_{m,i})
\eeq
is an injection.
\end{lem}
{\it Important remark.} The definition of the full set of the
Stokes' data requires a choice of the local coordinate in the neighborhood
of the puncture. But the data (\ref{data}) depend only on the $h_m$-jets
of the local coordinates, because it contains only diagonal matrices
$K_i$ with indices $i\leq 0$. We define an $h$-jet $[w]_{h}$
to be an equivalence class of $w$, with $w'$ and $w$ equivalent if
\beq\label{jet}
w'=w+O(w^{h+1}).
\eeq
{\it Proof.} Suppose that $\wt L$ and $\wt L_1$ have the same
data  (\ref{data}) modulo (\ref{trans}). Then solutions
$\Psi$ and $\Psi_1$ of the corresponding systems (\ref{sys}) have the same
monodromy along each cycle on $\G\backslash\{P_m\}$. Therefore,
$\phi=\Psi_1\Psi^{-1}$ is a single-valued meromorphic matrix function
on $\G\backslash \{P_m\}$. From (\ref{f3}) and (\ref{f6}) it follows that
$\phi$ is bounded in the neighborhood of $P_m$. Hence, $\phi$ is a
meromorphic function on $\G$ and is holomorphic at the points $P_m$.
The function $\Psi$ is  invertible everywhere except at the
the poles $\g_s$ of $\wt L$. The equation (\ref{sys}) implies that
vector rows of the residue of $\Psi_1\Psi^{-1}$ at $\g_s$ has the form
(\ref{F}). The assumption that $(\g,\a)$ are non-special Tyurin parameters
implies that $\phi$ is a constant matrix. Then, from the normalization
(\ref{n0}) it follows that $\Psi_1=\Psi$, and $\wt L=\wt L_1$.

Simple counting shows that $\A^D$ and the space of data (\ref{data}) modulo
transformations (\ref{trans}) have the same dimension. Therefore, the map
(\ref{data}) is a bijective correspondence of $\A^D$ and an open set
of the data.

\section{Isomonodromy deformations.}

Our next goal is to construct differential equations describing
deformations of $L\in \A^D(\G)$, which preserve a full set
of the data (\ref{data}). For brevity we call them isomonodromy
deformations. As it was mentioned above, in order to define the data
(\ref{data}) it is necessary to fix a normalization point $Q\in \G$,
a basis of $a_i, b_i$ cycles, paths $c_m$ connecting $Q$ with $P_m$,
and a set of $h_m$-jets of local coordinates in the neighborhoods of
the punctures $P_m$.

Let $h=\{h_m, \ \sum_m (h_m+1)=N\}$ be a set of nonnegative integers.
Then we denote the moduli space of smooth genus $g$ algebraic curves with
a puncture $Q\in \G$, and fixed $h_m$-jets of local coordinates $w_m$
in the neighborhoods of punctures $P_m$  by $\M_{g,1}(h)$.
The space $\A(h)$ of admissible meromorphic differentials
on algebraic curves with fixed multiplicities $(h_m+1)$ of the poles can be
seen as a total space of the bundle
\beq\label{abund}
\A(h)\ \longmapsto\  \M_{g,1}(h)=\{\G,P_m,[w_m], Q\}
\eeq
with fibers $\A^D(\G),\ D=\sum_m (h_m+1)P_m$. Here and below
$[w_m]$ stands for the $h_m$-jet of $w_m$. The space
$\M_{g,1}(h)$ is of dimension
\beq
\dim \ \M_g(h)=3g-2+N.
\eeq
An explicit form of the isomonodromy equations depends on a choice
of coordinates on $\M_{g,1}(h)$. Their Lax representation
requires in addition some sort of connection on the universal curve
$\N_g(h)$ which is a total space of the bundle
\beq\label{ucurve}
\N_g(h)\ \longmapsto \ \M_{g,1}(h)
\eeq
which fibers $\G$.

The following construction solves two problems simultaneously.
It goes back to a theory of the Whitham equations \cite{kr1,kr2}.
Details can be found in \cite{kp1,kp2}. First of all, locally we can
replace the moduli space of algebraic curves by the Teichmuller space
of marked algebraic curves, i.e. smooth algebraic curves with fixed basis
of $(a_i,b_i)$-cycles, and paths $c_m$ between $Q$ and punctures,
which do not intersect cycles.
Let us fix a set of integers $r_m,\ \sum_m r_m=0$. Then for any set of local
coordinates $w_m$ at $P_m$, there is a unique meromorphic differential $dE$
which in the neighborhood of $P_m$ has the form
\beq\label{e1}
dE=d\left(w_m^{-h_m}+r_m {\rm log}\,w_m+O(w_m)\right),
\eeq
and normalized by the condition
\beq\label{na}
\oint_{a_i}dE=0.
\eeq
The differential $dE$ depends only on the $h_m$-jets of the local coordinates
$w_m$. The zero divisor of $dE$ has degree $2g-2+N$. Let $\M_{g,1}^0(h)$ be
an open set of $\M_{g,1}(h)$ such that the corresponding differential $dE$
has simple zeros  $q_s\neq Q$
\beq\label{qs}
dE(q_k)=0,\ \ k=1,\ldots,2g-2+N.
\eeq
The Abelian integral
\beq\label{eQ}
E(q)=\int_{Q}^q dE,
\eeq
is single valued on the cover $\G^*$ of $\G\backslash \{P_m\}$ generated
by shifts along $b_i$-cycles and shifts along cycles $c_m'$ around
the punctures $P_m$. The curve $\G$ with cuts along $a_i$-cycles and
paths $c_m$ we regard as a marked sheet of $\G^*$.
The critical values
\beq\label{tt}
T_k=E(q_k),
\eeq
of $E$ on this sheet, and the $b$-periods of $dE$
\beq\label{tt1}
T_{b_i}=\oint_{b_i}dE,
\eeq
provide a system of local coordinates on $\M_{g,1}^0(h)$ (see details
in \cite{kp1}). The Abelian integral $E$ defines a local coordinate
on $\G^*$ everywhere except at the preimages $\wh q_s$ of the critical
points $q_s$. Therefore, $(E, T_k,T_{b_i})$ can be seen as a system of local
coordinates on an open set of a total space of the bundle $\N_g^*(h)$
over $\M_{g,1}(h)$ with fibers $\G^*$.

Let $\wt L(\tau)\in \A(h)$ be a one-parametric family of admissible
differentials. Its projection under (\ref{abund}) defines a path $T_a(\tau)$
in $\M_{g,1}(h)$. Here and below $\{a\}$ stands for the both types of
indices, i.e. $T_a=\{T_k,T_{b_i}\}$.
The family $\wt L(\tau)$ we regard as a family of one-forms
\beq\label{tcon}
\wt L(\tau)=L(E;T_a(\tau))dE,
\eeq
where $L$ is a function of the variable $E$ on $\G^*(\tau)$,
which is meromorphic everywhere except at $\wh q_s(\tau)$, and such that
\beq \label{c1}
L(E+T_{b_j}; T_a)=L(E; T_a),\ \ L(E+2\pi i r_m; T_a)=L(E; T_a).
\eeq
In the same way the corresponding solution $\Psi$ of equation (\ref{sys})
can be seen as a multi-valued function $\Psi(E; T_a)$ of the variable $E$
which is holomorphic everywhere except at $\wh q_s$ and the preimages
$\wh P_m$ on $\wh \G$ of the punctures $P_m$, and such that
\beq\label{c2}
\Psi(E+T_{b_j}; T_a)=\Psi(E; T_a)B_j,\ \
\Psi(E+2\pi i r_m; T_a)=\Psi(E; T_a) \mu_{m}.
\eeq
\begin{lem} A one parametric family of meromorphic connections
$\wt L(\tau)\in \A^{D(\tau)}_{\g(\tau),\a(\tau)}(\G(\tau))$ is an isomonodromy
family if and only if the logariphmic derivative of the corresponding
solution  $\Psi$ of (\ref{sys})
\beq\label{log}
M(E,\tau)=\p_{\tau} \Psi(E,\tau)\Psi^{-1}(E,\tau)
\eeq
as a function of $E$ is single valued on $\G^*(\tau)$,
(i) equals zero at $Q$, and is holomorphic everywhere except at the
points $\wh \g_s, \ \wh q_k$ where it has at most simple poles, (ii)
the vector rows of $M$ in the neighborhood
of $\wh \g_s$ have the form (\ref{F}),
(iii) the singular part of $M$ at $\wh q_k(\tau)$ equals
\beq\label{mk}
M(E,\tau)=-\p_{\tau}E(\wh q_k)\ L(E,\tau)+O(1),\
E\to E(\wh q_k) ,\ \
\eeq
(iv) $M$ satisfy the following monodromy properties
\begin{eqnarray}
M(E+T_{b_j}; T_a)&=&M(E; T_a)-(\p_{\tau}T_{b_j})\ L(E; T_a),
\label{c3} \\
M(E+2\pi i r_m; T_a)&=&M(E; T_a). \label{c4}
\end{eqnarray}
\end{lem}
{\it Proof.} The same arguments, as in the proof of the Lemma 3.2,
show that if the Stokes data do not depend on
$\tau$, then $M$ is holomorphic at the punctures
$\wh P_m$. The matrix $M$ is single-valued on $\G^*$ because monodromies
$A_j$ also do not depend on $\tau$. Unlike the previous case,
$M$ is single-valued only on $\G^*$, and acquires additional poles
at $\wh q_k$, because $E$ is multivalued on $\G$, and is not a local
coordinate at the critical points $\wh q_k$.

At the points $\wh q_k=\wh q_k(\tau)$ a local coordinate is
$(E-E(\wh q_k))^{1/2}$. Recall, that $E(\wh q_k)$ equals $T_k$ plus an integer
linear combination of $T_{b_j}$ which depends on the branch of $E$
corresponding to $\wh q_k$. The matrix function $\Psi$ is holomorphic
in the neighborhood of $\wh q_k$. Therefore, its expansion at $\wh q_k$
has the form
\beq\label{ps}
\Psi=\phi_0(\tau)+\phi_1(\tau)(E-E(\wh q_k))^{1/2}+O(E-E(\wh q_k)),
\wh q_k=\wh q_k(\tau).
\eeq
Then
\beq\label{ps1}
M=-\p_{\tau}E(\wh q_k)\
{\phi_1\phi_0^{-1}\over 2\sqrt{E-E(\wh q_k)}}+O(1).
\eeq
The logariphmic differential of $\Psi$ has the form
\beq\label{ps2}
d\Psi \Psi^{-1}=LdE={\phi_1\phi_0^{-1}dE\over 2\sqrt{E-E(\wh q_k)}}+O(1)dE.
\eeq
Equations (\ref{ps1}, \ref{ps2}) imply (\ref{mk}). Equations
(\ref{c3}, \ref{c4}) directly follow from (\ref{c2},\ref{sys}), and
the Lemma is proved.

Let us now introduce  basic functions $M_a$ corresponding to the
isomonodromy deformations along the coordinates $T_a$. Simple dimension
counting proves the following statement.
\begin{lem}  If $(\g,\a)$ is a non-special set of the Tyurin parameters,
then for each $\wt L\in \A^D_{\g,\a}(\G)$ there is a unique meromorphic
function $M_k$ on $\G$ such that: (i) $M_k$ is holomorphic everywhere
except at the points $\g_s$, and at the point $q_k$, (ii)
the vector rows of $M_k$ at $\g_s$ have the form (\ref{F}),
(iii) at the point $q_k$ the singular
part of $M_k$ has the form $M_k=-L+O(1)$, (iv) $M(Q)=0$.
\end{lem}
Let us denote $\G$ with a cut along the $a_i$ cycle by $\G_i^*$.
\begin{lem}  If $(\g,\a)$ is a non-special set of the Tyurin parameters,
then for each $\wt L\in \A^D_{\g,\a}(\G)$ there is a unique
function $M_{b_i}$ on $\G^*_i$ such that: (i) $M_{b_i}$ is holomorphic
everywhere except at the points $\g_s$, where the vector rows of $M_k$
have the form (\ref{F}), (ii) $M_{b_i}$ can be extended as a continuous
function on the closure of $\G_i^*$, and its boundary values $M_{b_i}^{\pm}$
on the two sides of the cut satisfy the relation
$M_{b_i}^+-M_{b_i}^- =-L$, (iii) $M_{b_i}(Q)=0$.
\end{lem}
A meromorphic matrix function on $\G^*_i$ which satisfies the boundary
condition (ii) can be represented as the Cauchy type integral over
the cycle. The difference of any two such functions is a
meromorphic function on $\G$. Therefore, once again a proof of the existence
and uniqueness of the function with prescribed analytical properties
is reduced to the Riemann-Roch theorem.

If we keep the same notations for pull back of $M_a$ on $\G^*$, then
the logariphimic derivative $M=\p_{\tau}\Psi \Psi^{-1}$ in the Lemma 4.1
can be written as
\beq\label{log1}
M=\sum_a (\p_{\tau} T_a)\  M_a.
\eeq
Now we are in position to define a hierarchy of differential equations
which describe the isomonodromy deformations. In the neighborhood of $\g_s$ the
Laurent expansion of $L=\wt L/dE$ and $M_a$ have the form
\begin{eqnarray}
L&=&{\b_s\a_s^T\over E-e_s}+L_{s1}+L_{s2}(E-e_s)+O((E-e_s)^2),
\ \ e_s=E(\g_s), \label{Lse}\\
M_a&=&{m_{s}^{a}\a^T_s\over E-e_s}+
M_{s1}^{a}+M_{s2}^{a}(E-e_s)+O((E-e_s)^2), \label{mse}
\end{eqnarray}
where $m_s^{a}$ are vectors.
\begin{th} The Lax equations
\beq\label{lax}
\p_a\wt L-dM_a+[\wt L,M_a]=0,
\eeq
define a hierarchy of commuting flows on $\A(h)$ which preserve
extended set of monodromy data (\ref{data}). They are equivalent to the
equations
\begin{eqnarray}
{}&\p_a e_s&=-\a_s^Tm_{s}^a, \ \ \ e_s=E(\g_s)\label{eq1}\\
{}&\p_a \a_s^T&=-\a_s^TM_{s1}^a-\lambda_s\a_s^T, \label{eq2}\\
{}&\p_a\b_s&=M_{s1}^a\b_s-\left(L_{s1}-\kappa_s\right)m_s^a+ \lambda_s\b_s,
\label{eq2a}\\
{}&\p_a\kappa_s&=\a_s^T\left(M_{s2}^a-L_{s2}\right)\b_s, \label{eq3}\\
{}&\p_a \wt L_m&=[M_a,\wt L_m]_{+}\ . \label{eq4}
\end{eqnarray}
where
$\lambda_s$ are scalar functions, and $[M_a,\wt L_m]_+$ denotes
the singular part of $[M_a,\wt L_m]$ at $P_m$.
\end{th}
Note that if $h_m=0$, then the right hand side of (\ref{eq4}) is just
$[M_a(q_k),\wt L_m]$.

\medskip
\noindent
{\it Proof.} First, let us show that the left hand side of (\ref{lax}),
which we denote by $\phi$, is a single-valued meromorphic function on
$\G$ which is holomorphic everywhere except at the points $\g_s$ and $P_m$.
Indeed, for $M_a=M_k$ it is single-valued by the definition of $M_k$,
but may have a pole at $q_k$. Taking derivative of the Laurent expansion of
$L$ at $q_k$, we obtain that $\p_a L$ acquires pole
at $q_k$ of the form $\p_a L=-dL/dE+O(1)$. Hence, the singular part of
$\p_a \wt L$ is just $-dL$, which cancels with the singular part of $dM_k$.
>From (\ref{eq4}) it follows that $[M_k,\wt L]$ is regular at $q_k$.
Almost identical arguments show that for $M_a=M_{b_i}$ the matrix
differentials $dM_a$ and $\p_a \wt L$ have the same monodromy
properties along the cycle $b_i$, and therefore, $\phi$ is single-valued
on $\G$.

Equations (\ref{eq1}-\ref{eq2a}) and (\ref{eq4}) are equivalent to the
condition that $\phi$ is a holomorphic matrix differential on $\G$.
Then equation (\ref{eq3}) is equivalent to the condition $\a_s^T\phi(\g_s)=0$.
That gives us a system of $r^2g$ linear equations for $\phi$.
As shown in \cite{kr0}, for non-special sets of the Tyurin parameters
these equation are linear independent, and therefore, imply $\phi=0$.

The matrix functions $M_a$ are uniquely defined by
$\wt L$. Hence equations (\ref{eq1}-\ref{eq4}) is a closed system
of differential equations on the space of parameters
$(e_s,\kappa_s,\a_s,\b_s,\wt L_m)$. Compatibility of the equations
for different indices $a$ is equivalent to the equation
\beq\label{ab}
\p_aM_b-\p_bM_a+[M_b,M_a]=0.
\eeq
In order to prove (\ref{ab}) we first check that the left hand side of
the equation is single valued meromorphic matrix function which
is holomorphic everywhere except at $\g_s$. Then, from equations
(\ref{eq1}-\ref{eq2a}) it follows that this function has at $\g_s$ at most
simple pole of the form (\ref{F}). For non-special sets of the Tyurin
parameters the last condition implies that the left hand side of
(\ref{ab}) is a constant matrix function on $\G$. It equals zero due
to normalization $M_a(0)=0$. Recall, that at the marked point $E(Q)=0$.

>From the Lax representation of equations (\ref{eq1}-\ref{eq4})
it follows that, if $\wt L$ is a solution of these equations
and$\Psi$ is the normalized solution of (\ref{sys}), then
$\p_a\Psi\Psi^{-1}=M_a$, . The Lemma 4.1 implies the
isomonodromy property of the flow, and the Theorem is proved.

\medskip
\noindent{\it Example 1.} The Schlesinger equations \cite{schles}
\begin{eqnarray}\label{s1}
\p_iA_j&=&{[A_i,A_j]\over t_i-t_j} ,\ \ i\neq j, \\
\p_iA_i&=&\sum_{j\neq i} {[A_j,A_i]\over t_i-t_j}\label{s1a}
\end{eqnarray}
describe the isomonodromy deformations of a meromorphic connection
with regular singularities
\beq\label{s2}
Ldz=\sum_i {A_i\over z-t_i}dz
\eeq
on the rational curve. In the conventional
approach the coordinates $t_i$ of the punctures on the complex
plane are considered as coordinates on the space of rational curve
with punctures. In our approach, which also works for higher
genus case, we use the function $E=\sum_i\ln(z-t_i)$ for
parameterization of points of the complex plane. Critical
values $T_k(t)$ of $E$
\beq\label{s3}
T_k=\sum_{i}\ln \ (q_k-t_i),
\eeq
locally define $t_i$ uniquely up to a common shift $t_i\to t_i+c$.
The critical points $q_k$ are roots of the the equation
\beq\label{s4}
E'(q_k)=\sum_i{1\over q_k-t_i}=0, \ \ E'(z)=\p_z
E(z).
\eeq
Note that (\ref{s4}) implies
\beq\label{s41}
E'(z)=\sum_i{1\over z-t_i}={\prod_k(z-q_k)\over \prod_i (z-t_i)}.
\eeq
>From (\ref{s4}) it follows that
\beq\label{s5} \p_i
T_k=-{1\over q_k-t_i}.
\eeq
According to the Lemma 4.2, the matrix
$M_k(z)$ corresponding to the variable $T_k$ has the only pole at
$q_k$ which coincides with the singular part of $-L/E_z$. Hence
\beq\label{s6}
M_k=-{\res_{q_k} (L/E')\over z-q_k}=-{1\over z-q_k}
\left(\sum_i{A_i\over q_k-t_i}\right) {\prod_j(q_k-t_j)\over
\prod_{s\neq k}(q_k-q_s)}
\eeq
>From equation (\ref{eq4}) we obtain
that ismonodromy deformations of $L$ with respect to new
coordinates have the form
\beq\label{s7}
\p_{T_k} A_j=-{1\over
t_j-q_k} \left(\sum_i{[A_i,A_j]\over q_k-t_i}\right)
{\prod_j(q_k-t_j)\over \prod_{s\neq k}(q_k-q_s)}
\eeq
It is instructive to check directly that equations (\ref{s7}) are
equivalent to (\ref{s1}). For $i\neq j$ we have
\beq\label{s8}
\p_iA_j=\sum_k(\p_iT_k)\p_{T_k}A_j=- \sum_k
\res_{z=q_k}{[L(z),A_j]\prod_{s\neq i,j}(z-t_s)\over
\prod_k(z-q_k)}
\eeq
The expression in right hand side of
(\ref{s8}) has poles at $t_i$ and $q_k$. Hence, $\p_iA_j$ equals
the residue of this expression at $z=t_i$.
\beq\label{s9}
\p_iA_j=[A_i,A_j]{\prod_{s\neq i,j}(t_i-t_s)\over
\prod_k(t_i-q_k)}
\eeq
Equation (\ref{s41}) implies
\beq\label{s10} 1=\res_{t_i}E'(z)={\prod_k(t_i-q_k)\over
\prod_{s\neq i}(t_i-t_s)}\ .
\eeq
Therefore, the last factor in
(\ref{s9}) equals $1/(t_i-t_j)$, and we obtain equation (\ref{s1}).
Equation (\ref{s1a}) can be replaced by the equation $\sum_i
\p_iA_j=0$. Therefore, it is enough to check that $\sum_i \p_iT_k=0$.
The last equation follows from (\ref{s4}), and (\ref{s5}).

\medskip
\noindent{\it Example 2.} The Painleve-II equation
\beq\label{p2}
u_{xx}-xu-2u^3=\nu
\eeq
describes an isomonodromy deformation of the rational connection
\beq\label{p3}
L=Az^2+Bz+C+Dz^{-1},\
\eeq
where
\beq\label{p3a}
A=-4i\s_3,\ B=-4u\s_2, \ C=-(2iu^2+x)\s_3-2u_x\s_1,\ D=\nu\s_2,
\eeq
and $\s_i$ are the Pauli matrices
\beq\label{p4}
\s_1=\left(\begin{array}{cc}0&1\\1&0\end{array}\right),\
\s_2=\left(\begin{array}{cc}0&-i\\i&0\end{array}\right),\
\s_3=\left(\begin{array}{cc}1&0\\0&-1\end{array}\right).
\eeq
We would like to stress once again, that the conventional definition of the
isomonodromy deformations of rational connections with irregular
singularities are ones which preserve  monodromy, transition, and the Stokes
matrices. Exponents $K_i$ are considered as parameters of the deformation
(see \cite{flashka, jimbo}).

In this example we show that the same equations can be seen as equations
describing deformations over the space of jets of local coordinate, which
preserve the full set of data (\ref{data}), including exponents.
Let us consider the isomonodromy deformation of $L$ corresponding to the
deformation of $z$ defined by the function
\beq\label{p5}
E(z)={4\over 3}z^3+xz+\ln z.
\eeq
The critical points $q_k$ are roots of the equation
\beq\label{p6}
4q_k^2+x+q_k^{-1}=0\longmapsto E'(z)=4z^{-1}\prod_{k=1}^3(z-q_k)
\eeq
As above, the Lax matrix $M_k$ corresponding to
the coordinate $T_k=E(q_k)$ equals
\beq\label{p7}
M_k=-{\res_{q_k} (L/E')\over z-q_k}
\eeq
As is (\ref{s5}), we obtain that $\p_xT_k=q_k$. Therefore,
if $\Psi$ is a solution of (\ref{sys}), then
\beq\label{p8}
\p_x\Psi(E)=M\Psi(E),\ \
M=-\sum_k {q_k\ \res_{q_k}(L/E')\over z-q_k}.
\eeq
In our notation we skip indication on an explicit dependence of functions
on $x$ but keep track of the variable which is considered fixed with respect to
$x$.

The matrix $M$ in (\ref{p8}) equals
\beq\label{p9}
M=-\sum_k\res_{q_k}F(q)=\res_{z}F(q)+\res_{\infty}F(q), \ \
F={qL(q)\over E'(q)(z-q)}\ .
\eeq
The residue at $q=z$ equals
\beq\label{p10}
\res_{\infty}F(q)=-{zL(z)/E'(z)}.
\eeq
Expansion of $F(q)$ at $q=\infty$ has the form
\beq\label{p11}
F=-{1\over 4}\left(Aq^2+Bq+C+Dq^{-1}\right)
\left(\sum_{s=0}^{\infty}zq^{-1}\right)
q^{-2}\left(1+O(q^{-2})\right).
\eeq
Therefore,
\beq\label{p12}
\res_{z}F(q)={1\over 4}(Az+B).
\eeq
Derivatives with fixed values of $E$ and $z$ are related to each other
by the chain rule
\beq\label{p13}
\p_x\Psi(x,z)=\p_x\Psi(x,E(z))+{d\Psi\over dE}\ \p_x E(x,z)=
\p_x\Psi(x,E(z))+{L(x,z)\over E'(x,z)}z
\eeq
Equations (\ref{p8}-\ref{p13}) imply
\beq\label{p14}
\p_x\Psi(x,z)=(Az+B)\Psi(x,z).
\eeq
The compatibility condition of (\ref{sys}) and (\ref{p14}) gives the
well-know Lax representation for (\ref{p2}).

\section{Hamiltonian approach}

In this section we show that the general algebraic approach to the Hamiltonian
theory of the Lax equations proposed in \cite{kp1,kp2,kr4}
is also applicable to the isomonodromy equations. Since the
arguments here are very close to the ones in the author's earlier work
\cite{kr0}, except for slight modifications , we shall be brief.

The entries of $\wt L\in \A(h)$ can be regarded as functions on
$\A(h)$ with values in the space of meromorphic differentials on $\G$.
Therefore, $\wt L$ by itself can be seen as matrix-valued function and
its external derivative $\delta \wt L$ as a matrix-valued one-form on $\A(h)$.
The formal solutions $\psi_m$ of the form (\ref{f3}) corresponding to
an expansion of $L$ at the punctures $P_m$ can also be regarded as
matrix function on $\A(h)$ defined modulo permutation of the columns and
the transformation
\beq\label{h1}
\psi_m'=\psi_m f_m ,
\eeq
where $f_m$ is a diagonal matrix. Hence, its differential $\delta \psi_m$ is
one-form on $\A^D$ with values in the space of formal series of the form
(\ref{f3}). In the same way we consider the differentials
$\delta K_i^{(m)}$ of the exponents in (\ref{f3}).

Let $\P_0$ be a subspace of $\A(h)$ such that restriction to
$\P_0$ of the differentials
$\delta K_i^{(m)}$ of the exponents in (\ref{f3}) vanishes for $i\leq 0$, i.e.
\beq\label{h2}
\delta K_i^{(m)}\big|_{\P_0}=0, \ \ i\le 0.
\eeq
In other words, $\P_0$ is a subspace of $\A(h)$ such that for $\wt L\in \P_0$
the singular parts $\wt L_m$ of $\wt L$ at the punctures are points of
a fixed set of orbits $\wt \O_m$ of the adjoint action of $GL_r^+(w)$ on
on the space of equivalence classes of meromorphic differentials at $P_m$,
modulo holomorphic differentials. Here $GL_r^+(w)$ is the group of invertible,
holomorphic in the neighborhood of $P_m$ matrix functions.

We define a scalar valued two-form on $\P_0$ by the the formula
\beq\label{form}
\omega=-{1\over 2}\left(\sum_{s=1}^{rg}\res_{\g_s} \wt \Omega+
\sum_{P_m}
\res_{P_m}\wt \Omega\right),
\eeq
where
\beq\label{om}
\wt \Omega=
{\rm Tr} \left(\psi^{-1} \delta \wt L\wedge \delta \psi\right),
\eeq
and $\psi$ in the neighborhood of $\g_s$ is a solution of (\ref{sys}),
and in the neighborhood of the puncture $\psi=\psi_m$ is the formal
solution (\ref{f3}).

Let us check that $\omega$ is well-defined. Indeed, if $\psi'=\psi g$
is another solution of (\ref{sys}) in the neighborhood of $\g_s$, then
\beq\label{h3}
\wt \Omega'=\wt \Omega+
{\rm Tr} \left[\left(\psi^{-1} \delta \wt L\psi\right)\wedge \delta g g^{-1}
\right].
\eeq
Taking external derivative of (\ref{sys}) we obtain the equalities
\beq\label{h4}
\delta d\psi=\delta \wt L \psi+\wt L\delta\psi,\
-\delta d\psi^{-1}=\psi^{-1}\delta \wt L+
\delta \psi^{-1}\wt L
\eeq
They imply
\beq\label{h5}
\psi^{-1} \delta \wt L\psi=d\left(\psi^{-1}\delta \psi\right)
\eeq
Therefore,
\beq\label{h55}
\wt \Omega'=\wt \Omega+{\rm Tr} \left[d(\psi^{-1}\delta \psi)\wedge
\delta g g^{-1}\right].
\eeq
The matrix $g$ is a constant matrix in the neighborhood of $\g_s$. Therefore,
the second term in (\ref{h3}) is a full differential of a meromorphic
function and does not contribute to the residue.

Consider now the residues of $\wt \Omega$ at the puncture $P_m$.
Essential singularities of $\psi_m$ and $\psi_m^{-1}$ cancel each other.
Therefore, $\wt \Omega$ in the neighborhood of $P_m$ is a formal meromorphic
differential, and its residue at $P_m$ is well-defined.
It does not depend on a permutation of columns of $\psi_m$.
Under the transformation (\ref{h1}) it gets additional term
\beq\label{h6}
{\rm Tr} \left[\left(\psi_m^{-1} \delta \wt L\psi_m\right)\wedge
\delta f_m f_m^{-1}\right]={\rm Tr} \left[
d\left(\psi_m^{-1}\delta \psi_m\right)\wedge \delta f_m f_m^{-1}\right].
\eeq
The matrix $f_m$ is diagonal, and therefore commutes with $K^{(m)}$.
The constraints (\ref{h2}) imply (\ref{h6}) is a holomorphic
differential, and therefore has zero residue.
\begin{th} The two-form $\omega$ defined by (\ref{form}) is gauge invariant
and descends to a closed, non-degenerate form on $\P=\P_0/SL_r$.
Under the correspondence (\ref{100}) it takes the form
\beq\label{tr3}
\omega=\sum_{s=1}^{rg}\left(\delta \kappa_s\wedge\delta z_s+\sum_{i=1}^r
\delta \b_s^i \wedge\delta \a_s^i\right)+\sum_m \omega_m
\eeq
where $\omega_m$ is the canonical symplectic structure
on an orbit $\wt \O_m$.

The isomonodromy equations (\ref{lax}) are Hamiltonian with respect
to symplectic structure defined by $\omega$. The Hamiltonians $H_a$ are equal
\beq\label{H}H_k=-{1\over 2}{\rm res}_{q_s}{\rm Tr}\left(\wt L^2/dE\right),\ \
H_{b_i}=-{1\over 2}\oint_{a_i} {\rm Tr}\left(\wt L^2/dE\right)
\eeq
\end{th}
Recall, that the tangent space to $\wt{\O}_m$ at
$\wt L_m$ is isomorphic to $sl_r^+/sl_r^+(\wt L_m)$, where $sl_r^+(\wt L_m)$
is the subalgebra of traceless matrix functions ${\xi}$ which are holomorphic
in the neighborhood of $P_m$, and such that
$[\wt L_m,{\xi}]$ is holomorphic at $P_m$. The symplectic structure on $\wt\O_m$
is defined by the formula (see details in \cite{kr0})
\beq\label{P4}
\omega_m=\res_{P_m} {\rm Tr} \left(\wt L_m\ [{\xi},{\eta}]\right).
\eeq
{\it Proof.} It is easy to check directly, that under the gauge
transformation
\beq\label{gau1}
L'=g^{-1}Lg,\ \psi'=g^{-1}\psi
\eeq
$\wt \Omega$ gets transformed to $\wt \Omega'=\wt \Omega+F$, where
\beq\label{trans1}
F={\rm Tr}\left(\psi^{-1}[\wt L,\delta h]\wedge \delta \psi-
[\wt L,\delta h]\wedge \delta h -\delta \wt L\wedge \delta h\right).
\eeq
Using (\ref{h4}), we obtain
\beq\label{trans2}
{\rm Tr}\left(\psi^{-1}[\wt L,\delta h]\wedge \delta \psi\right)=
{\rm Tr}\left(\delta h\wedge \delta \wt L-
d\left(\psi^{-1}\delta h\wedge \delta\psi\right)\right).
\eeq
The last term in (\ref{trans2}) is holomorphic in the neighborhoods
of $\g_s$ and $P_m$. The rest of $F$ is a global meromorphic differential
on $\G$ with the only poles at $\g_s$ and $P_m$. Therefore, sum of all
the residues of $F$ vanishes. Hence, $\omega$ is gauge invariant. Arguments
needed to complete a proof of (\ref{tr3}) are identical to the ones in
the proof of the Theorem 4.1 in \cite{kr0}.
>From (\ref{tr3}) it follows that $\omega$ is closed. It descends
to a non-degenerate form on $\P_m$, because $\omega_m$ is non-degenerate
on $\wt O_m$, and the first term in (\ref{tr3}) equals
\beq\label{tr55}
\omega_0=\sum_{s=1}^{rg}\delta \kappa_s\wedge\delta z_s+\sum_{s=r+1}^{rg}
\delta b_s^T \wedge \delta a_s, \ g>1,
\eeq
where $a_s,b_s$ are local coordinates on $\M$ defined by (\ref{pa34}).

Our next goal is to show that the isomonodromy equations are Hamiltonian
with respect to the symplectic form $\omega$. By definition a vector
field $\p_a$ on a symplectic manifold is Hamiltonian, if the contraction
$i_{\p_a}\omega(X)=\omega(X,\p_a)$ of the symplectic form
is an exact one-form $dH_a(X)$. The function $H_a$ is the Hamiltonian
corresponding to the vector field $\p_a$.

For each $\wt L\in \A^D_{\g,\a}$ let us define meromorphic differentials
$d\Omega_a=d\Omega_a(L)$. The differential $d\Omega_k$ is
a unique meromorphic differential on $\G$ that has pole of the form
\beq\label{h90}
d\Omega_k=-dL+0(1),\ \ \wt L=LdE,
\eeq
at $q_k$, and is holomorphic everywhere else, and satisfy the equations
\beq\label{h91}
\a_s^Td\Omega_a(\g_s)=0.
\eeq
The differential $d\Omega_{b_i}$ is a unique holomorphic differential on
$\G^*_i$ which satisfies (\ref{h91}), and is continuous on the
closure of $\G^*_i$. Its boundary values on two sides of the cut
along $a_i$-cycle satisfy the relation
\beq\label{h92}
d\Omega_{b_i}^+-d\Omega_{b_i}^-=-dL.
\eeq
\begin{lem} Evaluations of the one-forms $\delta L$ and $\delta \psi$ on the
vector field $\p_a$ defined by the Lax equation (\ref{lax}) equal
\begin{eqnarray}
\delta \wt L(\p_a)&=&\p_aL-d\Omega_a=[M_a,\wt L]+dM_a-d\Omega_a, \label{h7} \\
\delta \psi(\p_a)&=&M_a\psi+\phi_a, \label{h8}
\end{eqnarray}
where $\phi_a$ is a solution of the equation
\beq\label{h9}
d\phi_a=\wt L\phi_a-d\Omega_a\psi.
\eeq
\end{lem}
{\it Proof.} The right hand sides of (\ref{h7}, \ref{h8}) are not equal
to the derivatives of $\wt L$ and $\psi$, because by definition
$\delta$ is the external differential on a fiber $\A^D(\G)$, but not
on a total space of the bundle
$\A(h)$. In other words, if $I_k$ are coordinates on the space $\A^D(\G)$
on a fixed curve with punctures, then
\beq
\delta \wt L=\left(\p \wt L/\p {I_k}\right)\delta I_k\longmapsto
\delta \wt L(\p_a)=\left(\p \wt L/\p {I_k}\right)\p_aI_k.
\eeq
The data (\ref{100}) are coordinates on $\A^D(\G)$. From equations
(\ref{eq1}-\ref{eq4}) it follows
that the difference $\Phi$ of the both sides of (\ref{h7}) is a
holomorphic differential on $\G$ such that $\a_s^T\Phi(\g_s)=0$.
For non-special sets of the Tyurin parameters
the last equation implies $\Phi\equiv 0$.
Evaluation of (\ref{h4}) at $\p_a$, and equation (\ref{h7})
directly imply (\ref{h8}).

>From (\ref{h7}-\ref{h9}) it follows that the evaluation $\wt \Omega(\p_a)$
of the matrix valued two-form $\wt \Omega$ given by (\ref{om}) equals
\beq\label{h10}
\wt\Omega(\p_a)={\rm Tr} \left(
\psi^{-1}\delta L(M_a\psi+\phi_a)-\psi^{-1}[M_a,\wt L]\delta \psi\right).
\eeq
>From (\ref{h4}) and (\ref{h5}) it follows that
\beq\label{h11}
\wt \Omega(\p_a)={\rm Tr} \left(M_a\delta \wt L+\delta \wt LM_a-
\psi^{-1}d\Omega_a\delta \psi-
d(\psi^{-1}M_a\delta \psi)-d(\delta\psi^{-1}\phi_a)\right).
\eeq
The last two terms in (\ref{h11}) are differentials of meromorphic functions
in the neighborhoods of $\g_s$ and $P_m$. Therefore, their residues at
these points equal zero. From (\ref{h91}) it follows that the third
term is holomorphic at $\g_s$. It is holomorphic also at $P_m$.
For $M_k$ the first two terms are meromorphic
on $\G$ with poles at $\g_s,P_m$ and with a pole at the critical point
$q_k$. Hence,
\beq\label{h12}
i_{\p_a}\omega=
{1\over 2}\res_{q_k}{\rm Tr}\left(\delta \wt LM_k+M_k\delta \wt L\right).
\eeq
By definition, the matrix $M_k$ in the neighborhood of $q_k$ has the form
$M_k=-\wt L/dE+O(1)$. That implies (\ref{H}) for $T_a=T_k$. In the similar way
we prove that $\omega(\p_{T_{b_i}})$ equals to the external
differential of $H_{b_i}$, and therefore, the Theorem is proved.

The basic flows constructed above, easy alow to describe isomonodromy
equations,
corresponding to various subspaces of $\M_{g,1}(h)$, and to various
changes of coordinates. Let $T_a=T_a(\tau)$ depend on a variable $\tau$,
and let $z=z(E,T_a)$ be a local coordinate along $\G(T_a(\tau))$. Then the
matrix function $M$ which define a isomonodromic deformation of $\wt L$ in
the $\tau$-direction equals
\beq\label{h120}
M=\sum_{a}(\p_{\tau}T_a)M_a(z)+{\wt L\over dE}\p_{\tau}E(z)
\eeq
Let us consider the following instructive example.

\medskip
\noindent
{\bf Isomonodromy equations on a fixed algebraic curve}}.
A variation of the coordinates $T_a$ introduced above changes simultaneously
a curve, punctures and jets of local coordinates. In these coordinates
it is hard to identify variations that preserve $\G$.
For such deformations it is more convenient to use more traditional setting.

If $z=z_m$ be a local coordinate on $\G$ in an open domain $U_m$,
then the variables $t_m=z(P_m)$ are local coordinates on the space of
punctures $P_m\in U_m$.
Let $\wt L\in \A_{\g,\a}^D(\G)$ be an admissible meromorphic differential on
$\G$ with with regular singularities at $P_m$, i.e. in $U_m$ it has the form
\beq\label{hh12}
\wt L=\left({L_{m}\over z-t_m}+O(1)\right)dz,
\eeq
and corresponds to a non-special set of the Tyurin parameters $(\g,\a)$.

>From (\ref{h120}) it follows that $M^{(m)}$ corresponding to
the coordinate $t_m$ can be defined as a unique
meromorphic matrix function on $\G$ such that:
(i) $M^{(m)}$ is holomorphic on $\G$ everywhere
except at $\g_s$ and the point $P_m$; (ii) the rows of $M^{(m)}$ at $\g_s$
have the form (\ref{F}); (iii) in the neighborhood of $P_m$
the matrix $M^{(m)}$  has the form
\beq\label{h13}
M^{(m)}=-{L_m\over z-t_m}+O(1),
\eeq
and normalized by the condition $M^{(m)}(Q)=0$.
\begin{cor}
The Lax equations
\beq
\p_{t_m}\wt L-dM^{(m)}+[\wt L, M^{(m)}]=0,
\eeq
describe isomonodromy deformations of $\wt L$ with respect to the variables
$t_m$. They descend to the Hamiltonian equations on $\P$ with the Hamiltonians
\beq
H^{(m)}=-{1\over 2}\res_{P_m}{\rm Tr}\left(\wt L^2/dz\right).
\eeq
\end{cor}
A proof of the last statement is almost identical to that in the Theorem 5.1.
The differential $d\Omega_a$ in (\ref{h7}-\ref{h11}) has to be changed
to the differential $d\Omega^{(m)}$. The later
has pole the only pole at $P_m$, where
\beq\label{h130}
\Omega^{(m)}=-{L_m\over z-t_m}+0(1)
\eeq
It is normalized by the same condition (\ref{h91}).
As a result of that change the only term in (\ref{h11}) which
has nontrivial sum of the residues at $\g_s$ and $P_m$, is the third
term. It has nontrivial residue at $P_m$ which can be easy found using
(\ref{sys}).

\medskip
\noindent {\it Elliptic Schlesinger equations.} Let $Ldz$ be
a meromorphic connection on an elliptic curve $\G=C/\{2n\omega_1,2m\omega_2\}$
with simple poles at punctures $z=t_m$. In this example we denote the
parameters $\g_s$ and $\kappa_s$ by $q_s$ and $p_s$, respectively.

In the gauge $\a_s=e_s,\ e_s^j=\delta_s^j$ the $j$-th column of
the matrix $L^{ij}$ has poles only at the points $q_j$ and the punctures $t_m$.
Equation (\ref{Ls1}) implies  $L^{ji}(q_j)=0,\ i\neq j$. From equations
(\ref{Ls0},\ref{Ls1}) it follows that $L^{jj}$ at $q_j$ has the
expansion
$L^{jj}(z)=(z-q_j)^{-1}+p_j+O(z-q_j)$. An elliptic function with
these properties is uniquely defined by its residues $L_m^{ij}$ at the
punctures $t_m$, and can be written in terms of the Weierstrass
$\zeta$-function as follows
\begin{eqnarray}
L^{ii}(z)&=&p_i+
\sum_m L_m^{ii}\left( \zeta(z-t_m)-\zeta(z-q_i)-\zeta(q_i-t_m)\right)
, \ \ \sum_m L_m^{ii}=-1,\label{www1}\\
L^{ij}(z)&=&\sum_m L_m^{ii}\left( \zeta(z-t_m)-\zeta(z-q_j)-\zeta(q_i-t_m)+
\zeta(q_i-q_j)\right),\ \ i\neq j\label{www2}
\end{eqnarray}
The Poisson brackets are defined by the standard formulae
\beq\label{poi}
\{p_i,q_j\}=\delta_{ij},\ \{L_m^{ij},L_k^{ls}\}=\delta_{mk}
\left(-\delta_{jl}L_m^{is}+\delta_{is}L_m^{lj}\right)
\eeq
The elliptic Schelesinger equations are generated by the Hamiltonians
\begin{eqnarray}
H^{(m)}&=&-\sum_ip_iL_m^{ii}+\sum_i\sum_{k\neq m}L_m^{ii}L_k^{ii}
\left(\zeta(t_m-t_k)-\zeta(t_m-q_i)-\zeta(q_i-t_k)\right) \nonumber\\
&{}&-\sum_{i\neq j}L_m^{ij}L_m^{ji}\left(\zeta(q_j-q_i)-\zeta(t_m-q_i)-
\zeta(q_j-t_m)\right)\nonumber\\
&{}&-\sum_{k\neq m}\sum_{i\neq j}L_m^{ij} L_k^{ji}
\left(\zeta(t_m-t_k)-\zeta(t_m-q_i)-\zeta(q_j-t_k)-\zeta(q_i-q_j)\right).
\end{eqnarray}
\noindent{\it Example 3.} As an example of the isomonodromy equations
corresponding to deformations of algebraic curves, we consider
a meromorphic connection on an elliptic curve $\G=C/\{n,m\tau\}$
with one puncture, which without loss of generality we put at $z=0$.
That example in the framework of the Hamiltonian reduction approach was
considered in \cite{levin}.

We use the same gauge as in the previous example.
Let us assume that the residue of $\wt L$ at $z=0$ has the form
$-(1+h)+f$, where $1+h$ is a scalar matrix, and $f$ is a matrix of rank one:
$f^{ij}=a^ib^j$. As it was mentioned above, the equations $\a_i=e_i$ fix the
gauge up to
transformations by diagonal matrices. We can use these transformation to
make $a^i=b^i$. The corresponding momentum is given then by the collection
$(a^i)^2$ and we fix it to the values $(a^i)^2=h.$
Then, using the same arguments as before, we obtain that the matrix $L$
can be written as
\begin{eqnarray}
L^{ij}&=&h{\s(z+q_i-q_j)\, \s(z-q_i) \s(q_j) \over
\s(z)\s(z-q_j)\,\s(q_i-q_j)\,\s(q_i)}, \ i\neq j;\nonumber\\
L^{ii}&=&p_i+\zeta(z-q_i)-\zeta(z)+\zeta(q_i),
\end{eqnarray}
where $\s(z)=\s(z|1,\tau)$ is the Weierstrass $\s$-function.

According to the Theorem 5.1, the isomonodromy deformation of $L$ with respect
to the module $\tau$ of the elliptic curve is generated by the Hamiltonian
\beq
H=-{1\over 2}\int_0^1{\rm Tr}\  L^2dz
\eeq
The addition formula for the $\s$ function implies
\beq
\int_0^1L^{ij}L^{ji}dz=h^2\int_0^1(\wp(z)-\wp(q_i-q_j))dz=h^2(
2\eta_1-\wp(q_i-q_j)),
\ \ i\neq j.
\eeq
Here and below $\eta_1=\zeta(1/2),\eta_2=\zeta(\tau/2)$.
The formula
\beq
(\zeta(z-q_i)-\zeta(z)+\zeta(q_i))^2=\wp(z-q_i)+\wp(z)+\wp(q_i),
\eeq
and the monodromy property $\s(z+1)=-\s(z)e^{2\eta_1(z-1/2)}$ of the
$\s$-function imply
\beq
\int_0^1(L^{ii})dz=p_i^2+\wp(q_i)+2\eta_1+2p_i(\zeta(q_i)-2\eta_1q_i).
\eeq
The $(p,q)$-independent term in $H$ which is proportional to $\eta_1(\tau)$
does not effect the equations of motion. Therefore, the Hamiltonian
generating the isomonodromy equations for $p_i=p_i(\tau), q_i=q_i(\tau)$
equals
\beq\label{Ham}
-4\pi iH=\sum_i \left(p_n^2+2p_n(\zeta(q_n)-2\eta_1q_n)+\wp(q_n)\right)-h^2
\sum_{n\neq m}\wp(q_n-q_m).
\eeq
The equations of motion are
\begin{eqnarray}
q_{n,\tau}&=&-{1\over 2\pi i}(p_n+\zeta(q_n)-2\eta_1q_n),\label{cm1}\\
p_{n,\tau}&=&{1\over 4\pi i}\left(-2p_n(\wp(q_n)+2\eta_1)+\wp'(q_n)-
h^2\sum_{n\neq m}\wp'(q_n-q_m)\right).\label{cm2}
\end{eqnarray}
Equation (\ref{cm1}) implies
\beq\label{cm3}
q_{n,\tau\tau}=-{1\over 2\pi i}(\ p_{n,\tau}-
q_{n,\tau}(\wp(q_n)+2\eta_1)+\chi(q_n)),
\eeq
where
\beq\label{cm4}
\chi(z)=\chi(z;\tau)=\p_{\tau}\left(\zeta(z|1,\tau)-2\eta_1(\tau)z)\right)
\eeq
The function $\xi=\zeta(z|1,\tau)-2\eta_1(\tau)z$
has the following monodromy properties
$\xi(z+1)=\xi(z),\ \xi(z+\tau)=\xi(z)-2\pi i$. Therefore,
$\chi(z)$ is an entire function of $z$ such that
$\chi(z+1)=\chi(z),\ \chi(z+\tau)=\chi(z)-\p_z\xi(z)=
\chi(z)+(\wp(z)+2\eta_1)$. These analytic properties imply the following
expression for $\chi$ in terms of the Weierstrass functions:
\beq\label{cm5}
\chi(z)=
-{1\over 4\pi i}\left(2(\zeta(z)-2\eta_1z)(\wp(z)+2\eta_1)+\wp'(z)\right).
\eeq
>From (\ref{cm1}-\ref{cm5}) we get
\beq\label{cm6}
q_{n,\tau\tau}=-{h\over 8\pi^2}\sum_{n\neq m}\wp'(q_n-q_m|1,\tau).
\eeq
For $r=2$ equation (\ref{cm6}) for the variable $u=q_1-q_2$
is a particular case of the Painleve VI equation  (see details in
\cite{levin} and \cite{harnad}). It is to be said, that although equations
(\ref{cm6}) do coincide with one which were obtained in \cite{levin},
the Hamiltonian (\ref{Ham}) has intriguing new form.

\section{Canonical transformations}

In the previous section the symplectic form $\omega$, initially
defined by the formula (\ref{form}), was then expressed in terms of the
dynamical variables (\ref{100}). As a result it was identified
with the canonical symplectic structure on the space of  meromorphic
connections. The main goal of this section is to express $\omega$ in terms of
the monodromy data (\ref{data}).

Note first, that sum in (\ref{form}) is taken over all the poles of
$\wt L$. It is not equal to zero, because the solutions of (\ref{sys}), used
in (\ref{form},\ref{om}), are formal local solutions in the neighborhoods of
the punctures. Consider now the differential $\Omega_0$
given by  same formula, as $\wt \Omega$ in (\ref{om}), i.e.
\beq\label{ct0}
\Omega_0=
{\rm Tr} \left(\Psi^{-1} \delta \wt L\wedge \delta \Psi\right),
\eeq
but where $\Psi$ is a (global) multi-valued holomorphic solution of
(\ref{sys}) on $\G\backslash \{P_m\}$.
The differential $\Omega_0$ is single-valued on $\G$ with cuts along
$(a_k,b_k)$-cycles and paths $c_m$  between the marked point $Q$ and the
punctures $P_m$. Therefore,
\beq\label{ct1}
\sum_{s=1}^{rg}\res_{\g_s} \Omega_0={1\over 2\pi i}\oint_{\L}\Omega_0-
{1\over 2\pi i}\oint_{\C}\Omega_0,
\eeq
where $\L=\prod_{k=1}^g (a_kb_ka_k^{-1}b_k^{-1})$, and $\C=\prod_m C_m$
are loops in $\G\backslash \{P_m\}$ (see fig.1).

$$
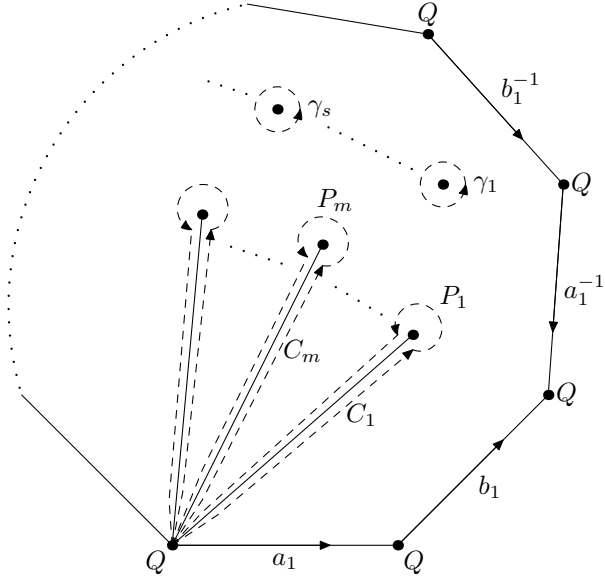
\epsfbox{figure1.ps}$$

\noindent
If $\Psi(Q)=1$ at the initial point, then the monodromy of $\Psi$ along the
loop $aba^{-1}b^{-1}$ is equal to
\beq\label{ct1a}
J(A,B)=B^{-1}A^{-1}BA,
\eeq
where $A,B$ are the monodromies corresponding to $a$- and $b$-cycles.
The monodromy of $\Psi$ along $b$ segment of the loop is $A^{-1}BA=BJ$.
>From (\ref{h55}) it follows that the sum of integrals of $\Omega_0$
along the $a$ and $a^{-1}$ segments of the loop is equal to
\begin{eqnarray}\label{ct2}
I_1&=&
-{\rm Tr} \left(A^{-1}\delta A \wedge \delta (BJ) J^{-1}B^{-1}\right)
\nonumber\\
{}&=&-{\rm Tr} \left[A^{-1}\delta A \wedge \delta B B^{-1}+
B^{-1}A^{-1}(\delta A)B \wedge \delta J J^{-1}\right]\, .
\end{eqnarray}
The monodromy of $\Psi$ along the $a^{-1}$-segment of the loop is
$A^{-1}J$. Therefore, sum of the integrals of
$\Omega_0$ along $b$ and $b^{-1}$ segments of the loop is equal to
\begin{eqnarray}\label{ct4}
I_2&=&-{\rm Tr} \left[(A^{-1}B^{-1}\delta (B A)-A^{-1}\delta A)\wedge
\delta (A^{-1}J)J^{-1}A\right]\nonumber\\
{}&=&-{\rm Tr}\left[-B^{-1}\delta B\wedge \delta A A^{-1}+
B^{-1}\delta B\wedge \delta J J^{-1}\right]\, .
\end{eqnarray}
The sum $\chi=I_1+I_2$ equals
\beq
\label{ct4a}
\chi(A,B)={\rm Tr} \left[B^{-1}\delta B\wedge \delta A A^{-1}-
A^{-1}\delta A \wedge \delta B B^{-1}+\delta J J^{-1}\wedge
B^{-1}A^{-1}\delta (AB) \right] .
\eeq
Due to analytical continuation, the solution $\Psi$ on the segment
of the loop $\L$ differs from the normalized solution $\Psi_0$ used
in the previous formulae by the factor
\beq\label{ct5}
H_1=1;\  H_k=J_{k-1}J_{k-2}\cdots J_1,\ k>1;
\  \ J_s=J(A_s,B_s).
\eeq
>From (\ref{h55}) it follows that the integral
of $\Omega_0$ over $(a_kb_ka_k^{-1}b_k^{-1})$ under the
transformation $\Psi= \Psi_0 H_k$
gets additional term
\beq\label{ct6}
{\rm Tr}\left(J_k^{-1}\delta J_k\wedge \delta H_k H_k^{-1}\right).
\eeq
Let us denote the integral of $\Omega_0$ over $\L$ by
\beq\label{ct7}
\omega_1({\bf A},{\bf B}):=
\oint_{\L} \Omega_0=\sum_{k=1}^g \left[\chi(A_k,B_k)+
{\rm Tr}\ \left(J_k^{-1}\delta J_k\wedge \delta H_k H_k^{-1}\right)\right].
\eeq
It is a two-form on the space of sets of matrices ${\bf A}=\{A_k\},\ {\bf B}=
\{B_k\}$.

Next we compute the integral of $\Omega_0$ along the cycle
$C_m$, which goes along one side of the cut $c_m$, then
goes around $P_m$ along a small circle $c_m'$,  and finally goes back
along the other side of $c_m$ (see fig.1).

Consider first the integral of $\Omega_0$ around the puncture. We split
the circle $c'$ into $2h+1$ arcs $c_{\nu}$ which lies in the sectors $V_{\nu}$
(here and below we skip for brevity the index $m$ of the puncture).
Recall, that in each of the sectors the formal solution $\psi$ of (\ref{sys})
given by the Lemma 3.1 is an asymptotic series for the holomorphic function
$\Psi_{\nu}=\Psi g_{\nu}^{-1}$. Let $\Omega_{\nu}$ be given by the same
formula as for $\Omega_0$ with $\Psi$ replaced by $\Psi_{\nu}$. Then,
\beq\label{ct10}
\int_{c_{\nu}}\Omega_0=\int_{c_{\nu}}\Omega_{\nu}+{\rm Tr}\left[\int_{c_{\nu}}
d\left(\Psi_{\nu}^{-1}\delta \Psi_{\nu}\right)\wedge
\delta g_{\nu}g_{\nu}^{-1}\right]
\eeq
The form $\wt \Omega$ defined by (\ref{om}), where $\psi$ is the formal
solution (\ref{sys}) gives an asymptotic series for $\Omega_{\nu}$ in
$V_{\nu}$. Therefore, as $c'$ shrinks to the puncture
\beq\label{ct11}
\lim_{c'\to P} \ \sum_{\nu}\int_{c_{\nu}}\Omega_{\nu}=(2\pi i)\
\res_P\  \wt\Omega.
\eeq
A sum of the second terms in (\ref{ct10}) equals
\begin{eqnarray}\label{ct12}
&{}&{\rm Tr}\left(\Psi_{2h+1}^{-1}(p)\delta \Psi_{2h+1}(p)\wedge
\delta g_{2h+1}g_{2h+1}^{-1}-\Psi_{1}^{-1}(p)\delta \Psi_{1}(p)\wedge
\delta g_{1}g_{1}^{-1}\right)+ \nonumber\\
&{}&\sum_{\nu=1}^{2h}{\rm Tr}
\left(\Psi_{\nu}^{-1}(p_{\nu})\delta \Psi_{\nu}(p_{\nu})\wedge
\delta g_{\nu}g_{\nu}^{-1}-
\Psi_{\nu+1}^{-1}(p_{\nu})\delta \Psi_{\nu+1}(p_{\nu})\wedge
\delta g_{\nu+1}g_{\nu+1}^{-1}\right),
\end{eqnarray}
where $p_{\nu}\in V_{\nu}\cap V_{\nu+1}$ is common edge point of the arcs
$c_{\nu}$ and $c_{\nu+1}$. The point $p$ is the intersection point of the cut
$c$ and the circle $c'$. We assume that the cut tends to the puncture
in the intersection $V_1\cap V_{2h+1}$.

The matrices $\Psi_{2h+1}$ and $\Psi_1$ are connected by the relation
$\Psi_{2h+1}=\Psi_1e^{2\pi iK_0}$. Recall, that
the monodromy $\mu$ along the whole path $C$ is
$\mu=g_1^{-1}e^{2\pi i K_0}g_{2h+1}$.
Therefore, the first two terms in (\ref{ct12}) give
\beq\label{ct14}
{\rm Tr}\ \left[\Psi_1^{-1}\delta \Psi_1\wedge\left(
e^{2\pi iK_0}\delta g_{2h+1}g_{2h+1}e^{-2\pi iK_0}-\delta g_1 g_1^{-1}\right)
\right]={\rm Tr}\
\left(\Psi_1^{-1}\delta \Psi_1\wedge g_1\delta \mu \mu^{-1}
g_1^{-1}\right).
\eeq
Boundary values of $\Psi$ on the two sides of the cut $c$ between $Q$ and $P$
satisfy the relation $\Psi^+=\Psi^- \mu$. Therefore, sum of the integrals of
$\Omega_0$ along the first and the last segments of the path $C$ equals
\beq\label{ct9}
-{\rm Tr} \left(\Psi^{-1}(p)\delta \Psi(p \wedge \delta \mu \mu ^{-1}\right)=
{\rm Tr} \left(\delta \mu  \mu^{-1}\wedge g_1^{-1}\delta g_1+
\delta \mu \mu^{-1}\wedge g_1^{-1}\Psi_1^{-1}(p)\delta \Psi_1(p)g_1\right),
\eeq
Here we use the relation $\Psi(p)=\Psi_1(p)g_1$.
The sum of (\ref{ct14}) and (\ref{ct9}) is equal to
\beq\label{ct17}
I_3= {\rm Tr}\left(\delta \mu \mu^{-1}\wedge g_1^{-1}\delta
g_1\right).
\eeq
Recall, that $\Psi_{\nu+1}=\Psi_{\nu}S_{\nu}$ where the Stokes
matrix $S_{\nu}$ equals $S_{\nu}=g_{\nu}g_{\nu+1}^{-1}$.
Therefore, the terms of the sum in (\ref{ct12}) are equal to
$$
{\rm Tr}\left[-S_{\nu}^{-1}\delta S_{\nu}\wedge \delta g_{\nu+1} g_{\nu+1}^{-1}
+\Psi_{\nu}^{-1}(p_{\nu})\delta \Psi_{\nu}(p_{\nu})\wedge
\left(\delta g_{\nu} g^{-1}_{\nu}-S_{\nu}\delta g_{\nu+1} g_{\nu+1}^{-1}
S_{\nu}^{-1}\right)\right]=
$$
$$
={\rm Tr}\left(-\delta S_{\nu}S_{\nu}^{-1}\wedge \delta
g_{\nu} g_{\nu}^{-1}+\Psi_{\nu}^{-1}\delta \Psi_{\nu}\wedge
\delta S_{\nu}S_{\nu}^{-1}\right).
$$
In the sector $V_{\nu}$ we have $\Psi_{\nu}e^{-Kw^{-h}}=O(1)$, where
$K$ is the is the leading exponent $K_{-h}$. Therefore, (\ref{S1})
implies
\beq
\lim_{p_{\nu}\to P}{\rm Tr}\left(\Psi_{\nu}^{-1}\delta \Psi_{\nu}\wedge
\delta S_{\nu}S_{\nu}^{-1}\right)=0.
\eeq
Hence, the second term in (\ref{ct12}) tends to
\beq\label{ct20}
I_4=\sum_{\nu=1}^{2h}{\rm Tr} \left(-\delta
S_{\nu}S_{\nu}^{-1}\wedge \delta g_{\nu}g_{\nu}^{-1}\right),
\eeq
as $c'$ shrinks to $P$.
Let us denote the sum of (\ref{ct17}) and (\ref{ct20}) by
\beq\label{ct18b}
\s({S},G,K_0)=
{\rm Tr}\left(\delta \mu \mu^{-1}\wedge G^{-1}\delta
G-\sum_{\nu=1}^{2h}\delta S_{\nu}S_{\nu}^{-1}\wedge
\delta g_{\nu}g_{\nu}^{-1}\right),
\eeq
where $S=\{S_{\nu}\}$, and matrices $\mu, g_{\nu}$
are given by (\ref{S5}, \ref{f6}). The integral of $\Omega_0$ along $C$
equals $\s(S,G,K_0)$ under the assumption that
$\Psi=\Psi_0$, where $\Psi_0=1$ at the initial point of the cycle.
Due to analytical continuation along the path $C=\prod_{m}{C_m}$
the initial value for the cycle $C_m$ equals
\beq\label{ct18}
F_1=1;\ F_m=\mu_{m-1}\mu_{m-2}\cdots \mu_{1}, \ m>1.
\eeq
>From (\ref{h55}) it follows that
the integral of $\Omega_0$ along the segment $C_m$ of the path $C$
under the transformation $\Psi=\Psi_0F_m$ acquires additional term
\beq\label{ct18a}
{\rm Tr}\ \left(\mu_{m}^{-1}\delta \mu_{m}\wedge \delta F_m F_m^{-1}
\right).
\eeq
Let us define a family of two-forms on the space of sets of
matrices ${\bf S}=\{S_{\nu}^{(m)}\},{\bf G}=\{ G^{(m)}\}$
parametrized by a set ${\bf K_0}=\{K_0^{(m)}\}$ of diagonal matrices:
\begin{eqnarray}\label{ct21}
\omega_2({\bf S},{\bf G}\, |\, {\bf K_0})&:=&
\oint_{\C} \Omega_0-2\pi i\ \sum_{m}\res_{P_m}\wt \Omega\nonumber\\
{}&=&\sum_m\left[\
\s(S^{(m)},G_m,K_0^{(m)})+
{\rm Tr} \left(\mu_{m}^{-1}\delta \mu_{m}\wedge \delta F_m
F_m^{-1}\right)\right].
\end{eqnarray}
Summarizing we obtain the following statement.
\begin{th}
The symplectic form $\omega$ defined by (\ref{form}) is equal to
\beq\label{ct23}
\omega={1\over 4\pi i}\ \left[\ \omega_2({\bf S}, {\bf G}\, |\,  {\bf K_0})-
\omega_1 ({\bf A},{\bf B})\ \right],
\eeq
where ${\omega_1}$ and ${\omega_2}$ are given by
(\ref{ct7}) and (\ref{ct21}), respectively.
\end{th}
It would be quite interesting to check directly that the formula
(\ref{ct23}) defines a symplectic structure on orbits of the adjoint action
of $SL_r$ on the space of  the  sets $ ({\bf A},{\bf B},{\bf S},{\bf G})$,
of matrices, which satisfy the only relation
\beq\label{ct30}
\prod_k^{\leftarrow}(B_k^{-1}A_k^{-1}B_kA_k)=\prod_m^{\leftarrow}\mu_{m}.
\eeq
The factors in (\ref{ct30}) are ordered such that indices increase from
right to left.

\medskip
\noindent{\it Example.} Let us consider the case of meromorphic connections
on the rational curve with one irregular singularity of order 2 and one regular
singularity. Without loss of generality we assume that $\wt L=Ldz$
has irregular singularity at $z=0$ and regular singularity at $z=\infty$, i.e.
\beq\label{b1}
L=l_{-1}z^{-2}+l_0z^{-1}.
\eeq
Let us fix a gauge in which $l_1=K_0^{\infty}$ is a diagonal matrix.
Then the monodromy matrix at the infinity is
$\mu_{\infty}=\exp (2\pi i K_0^{\infty})$. Recall,
that we always assume that the exponents are fixed.
The monodromy data at $z=0$ are two Stokes matrices $S_1,S_2$, transition
matrix $G$, and the exponents $K_1, K_0$. The monodromy matrix at $z=0$
equals
\beq\label{b2}
\mu_0=G^{-1}e^{2\pi K_0}S_2^{-1}S_1^{-1}G=\mu_{\infty}^{-1}\ .
\eeq
The transition matrices to the first and the second sectors at $z=0$
equal
\beq\label{b3}
g_1=G,\ \ g_2=S_1^{-1}G
\eeq
Substitution of (\ref{b3}) in (\ref{ct18b}) implies
\beq\label{b4}
4\pi i\ \omega=-{\rm Tr}\left(\delta S_1S_1^{-1}\wedge \delta G G^{-1}+
\delta S_2 S_2^{-1}\wedge \delta (S_1^{-1}G)G^{-1}S_1\right) .
\eeq
Using skew-symmetry of the wedge product and (\ref{b2}) we can rewright
the last term as
\begin{eqnarray}\label{b5}
{\rm Tr}\left(\delta S_2 S_2^{-1}\wedge \delta (S_1^{-1}G)G^{-1}S_1\right)&=&
{\rm Tr}\left(S_2^{-1}\delta S_2 \wedge \delta
(S_2^{-1}S_1^{-1}G)G^{-1}S_1S_2\right)=\nonumber \\
{}&=&{\rm Tr}\left(e^{2\pi i K_0}S_2^{-1}\delta S_2 e^{-2\pi iK_0}\wedge
\delta G G^{-1}\right).
\end{eqnarray}
Hence,
\beq\label{b6}
\omega=-{1\over 4\pi i}{\rm Tr}\left[\left(\delta S_1S_1^{-1}+
e^{2\pi i K_0}S_2^{-1}\delta S_2 e^{-2\pi iK_0}\right)\wedge
\delta G G^{-1}\right].
\eeq
The formula (\ref{b6})
after change of notations $G=C, S_1=b_+^{-1}, S_2e^{-2\pi K_0}=b_-$
coincides (up to a factor 2) with the formula (14) in \cite{boalch1},
where the symplectic structure on the space of monodromy data for
the linear system (\ref{b1}) was idenitfied with the sumplectic structure
of the group $G^*$ dual to $G=GL_r$.

\section{The Whitham equations}

It is well-known that the family of flat $\e\neq 0$-connections on
holomorphic vector bundles over an algebraic curve $\G$ with punctures extends
to a smooth family over the whole $\e$-plane. The central fiber over $\e=0$
is identified with the cotangent bundle to the moduli space of holomorphic
vector bundles on $\G$. The correspondence (\ref{100}) makes these
statements transparent.

The space of meromorphic $\e$-connections with fixed multiplicities $h=\{h_m\}$
of poles is the factor space $\A_{\e}(h)/SL_r$ of the space
of meromorphic differentials $\wt L_{\e}$ such that $\e^{-1}L_{\e}\in \A(h)$.
A meromorphic differential $\wt L_{\e}\in \A_{\e}(h)$ satisfies the constraints
(\ref{Ls},\ref{Ls0},\ref{Ls1}), and the condition
\beq\label{w}
\res_{\g_s}{\rm Tr}\ \wt L=(\a_s^T\b_s)=\e.
\eeq
The characteristic property of meromorphic
$\e$-connections is that in the neighborhood of the points $\g_s$
they have the form
\beq\label{wgauge}
\wt L=\e d\Phi_s(z)\Phi_s^{-1}(z)+\Phi_s(z)\wt L_s(z)\Phi_s^{-1}(z),
\eeq
where $\wt L_s$ and $\Phi_s$ are holomorphic at $\g_s$, and $\det \Psi_s$
has at most simple zero at $\g_s$.
In the earlier work of the author \cite{kr0} the space $\A_0(h)$ was called
the space of Lax matrices, and orbits of the adjoint action of $SL_r$ on
subspaces of $\A_0(h)$ with fixed singular parts of the eigenvalues were
identified with phase spaces of the Hitchin systems.

The space $\A_{\e}(h)$ is a bundle over the moduli space
$\M_{g,1}(h)$. Let $\wt L_{\e}(\tau)$ be a one-parametric deformation of
$\wt L_{\e}$ which preserves the full set of monodromy
data (\ref{data}) associated with a holomorphic solution of the equation
\beq\label{esys}
\e d\Psi=\wt L\Psi,
\eeq
Along identical lines to the proof of Lemma 4.1 it can be shown
that the singularities of $M=\p_{\tau}\Psi \Psi^{-1}$ are of the form
$\e^{-1}\wt L_{\e}/dE$. Therefore, in order to get a smooth at $\e=0$ family
of the isomonodromy equations, it is necessary to make a proper rescaling
of coordinates on $\M_{g,1}(h)$. Namely, if we introduce the {\it fast}
coordinates $t_a=e^{-1}T_a$, then the isomonodromy equations are equivalent
to the Lax equations
\beq\label{elax}
\p_{t_a}\wt L_{\e}-\e dM_a+[\wt L_{\e},M_a]=0,
\eeq
where matrices $M_a=M_a(\wt L_{\e})$ are defined by the same analytical
properties as above in Section 4. Moreover, the corresponding
Hamiltonians are given by the same formulae (\ref{H}).

\noindent{\it Remark.} Here and below we use the coordinates
$T_a$ on $\M_{g,1}$ introduced in Section 4, but mainly our arguments
do not rely on any specific choice of the coordinates.

As follows from \cite{kr0}, equations (\ref{elax}) for $\e=0$
\beq\label{olax}
\p_{t_a}\wt L_0=[M_a,\wt L_0]
\eeq
coincide with the Lax equations for commuting flows of the Hitchin system
corresponding to the second order Hamiltonians given by the same formula
(\ref{H}). Equations (\ref{olax}) describe {\it isospectral} deformations.
If $\wt L_0\in \A_0(h)$ is a solution of (\ref{olax}), then the spectral
curve $\wh \G$ of $\wt L_0$ defined by the characteristic equation
\beq\label{scurve}
\det(\wt k-\wt L_0)=\wt k^r+\sum_i u_i \wt k^i=0,
\eeq
is {\it time-independent}. The spectral transform identifies $\A_0(h)$ with
the Jacobian bundle over the moduli space $\SP$ of the spectral curves.
The fiber of this bundle over $\wh \G$ is the Jacobian $J(\wh \G)$.
The bijective correspondence between $\A(h)$ and the
Jacobian bundle over $\SP$ can be seen as a parametrization
of $\A_0(h)$ in the form $\wt L_0=\wt L_0(\phi|I)$. Here and below
we regard $\wt L_0(\phi|I)$ as an abelian function of the variable
$\phi\in J(\wh \G)$ depending on $I\in \SP$. The function $\wt L_0(\phi|I)$
takes values in the space of meromorphic matrix differentials on $\G$.
The equations of motions are linearized on the Jacobian of the spectral
curve, and therefore, the general solution of (\ref{olax}) can be represented
in the form $\wt L_0=\wt L_0(Ut|I)$, where
$Ut=\sum_aU_at_a$, and $U_a=U_a(I)$ are constant vectors depending on $I$
(see details in \cite{kr0}).

The main goal of this section is to apply ideas of the Whitham
averaging method to construct asymptotic solutions of the isomonodromy
equations (\ref{elax})
\beq\label{w20}
\wt L_{\e}=\wt L_0+\e \wt L_1+\e^2 \wt L_2+\cdots,\
M_{\e}=M_0+\e M_1+\e^2M_2+\cdots,
\eeq
where the leading terms have the form
\beq\label{w21}
\wt L_0(\e^{-1}S(T)|I(T)), \ M_0=(\e^{-1}S(T)|I(T)),
\eeq
and $T=\e t$ are {\it slow} variables.
If the vector-function function $S(T)$ satisfies the equation
\beq\label{w22}
\p_T S(T)=U(I(T))=U(T), \ \ {\rm i.e.} \ \
S(T)=\int^T U(T)dT,
\eeq
then  the leading term of (\ref{w20})
satisfies the original equation up to first order one in $\e$. All
the other terms of the asymptotic series are obtained
from the non-homogeneous linear equations with a homogeneous part
which is just the linearization of the original non-linear
equation on the background of the exact solution $\wt L_0$. In
general, the asymptotic series becomes unreliable on scales of the
original variables $t$ of order $\e^{-1}$. In order to have a
reliable approximation, one needs to require a special dependence
of the parameters $I(T)$. Geometrically, we note that
$\e^{-1}S(T)$ agrees to first order with $Ut$, and $t$ is the fast
variables. Thus $\wt L_0(\e^{-1}S(T)|I(T))$ describes a motion which
is to first order the original {\it fast} periodic motion on the
Jacobian, combined with a slow drift on the moduli space of exact
solutions. The equations which describe this drift are in general
called {\it Whitham equations}, although there is no systematic
scheme to obtain them.

Below we follow  lines of the scheme proposed in \cite{kr1},
where the Whitham equations for general $(2+1)$ integrable soliton systems
were derived.
First, we introduce sets of Abelian differentials $dv^r_a,dv^{i}_a$ on
spectral curves $\wh \G$.
The differentials $dv^r_k,dv^i_k$ are {\it real normalized}, i.e. their
periods are pure imaginary
\beq
{\it Re }\oint_c dv^r_a={\rm Re} \oint_c dv^i_a=0,\
c\in H^1(\wh\G).
\eeq
For indices $a=k$ the corresponding differentials have pole only at the
preimages $q_k^j$ on $\wh \G$ of the point $q_k$, where
\beq
dv_k^r=dk_j(z)+O(1),dv^i_k=idk_j(z)+O(1),
\eeq
and $k_j(z)$ is the corresponding branch of the eigenvalue of $\wt L(z)/dE$,
i.e. the corresponding root of the equation
\beq\label{scurve1}
\det(k-\wt L_0/dE)=0.
\eeq
For indices $a=b_j$ the corresponding differentials are holomorphic on
$\wh \G$, with cuts along all the preimages $a_j^l\in H^1(\wh \G)$ of the
cycle $a_j\in H^1(\G)$, and their boundary values on two sides of the cut
$a_j^l$ satisfy the relation
\beq
(dv^r_{b_j})^+-(dv^r_{b_j})^-=dk_l,\ \ (dv^i_{b_j})^+-(dv^i_{b_j})^-=idk_l
\eeq
Here, as before, $k_l$ is the corresponding eigenvalue of $\wt L/dE$.
Note, that $dv^r_a+idv^i_a$ is a holomorphic differential.
\begin{th} A necessary condition for the existence of the asymptotic
solution (\ref{w20}) of the isomonodromy equation (\ref{elax}) with the
leading term (\ref{w21}) and with bounded first correction term $\wt L_1$
is the equations
\beq\label{W}
\p_{X_a}\wt k=-dv^r_a,\ \ \p_{Y_a}\wt k=-dv^i_a.
\eeq
where $X_a$ and $Y_a$ are the real and the imaginary parts of the slow
variable $T_a=X_a+iY_a$.
\end{th}
Along lines of (\cite{kr1}) it can be shown that equations (\ref{W})
are generating form of the equations on the space $\SP$ of the spectral
curves (see details in \cite{kp1,kp2,kr2,kr5}).

Equations (\ref{W}) can be written in the form
\beq\label{W2}
\p_{T_a} \wt k=-dv_a,
\eeq
where
\beq
\p_{T_a}={1\over 2}\left({\p\over \p x_a}-i{\p\over\p y}\right),\ \
dv_a={1\over 2}(dv^r_a-idv^i_a).
\eeq
{\it Remark.} The equation (\ref{W2}) is a particular case of the exact
solutions of the universal Whitham hierarchy. It is connected with a theory
of WDVV equations and the Seiberg-Witten theory of $N=2$ supersymmetric
gauge models (see \cite{kp1,kp2,kr2,kr5}).
\begin{cor}
The real parts of the periods of the differential $\wt k$ over
the spectral curve are integrals of the Whitham equations.
The correspondence
\beq
\wt L_0\longmapsto  \oint_c K,\ c\in H^1(\wh \G)
\eeq
defines a flat connection on the bundle $\SP$ over $\M_{g}(h)$
\end{cor}
{\it Proof of the Theorem.}
Substitution of the series (\ref{w20}) into (\ref{elax}) gives
non-homogeneous linear equation for the first order terms
\beq\label{i1}
\p_t \wt L_1-dM_1+[\wt L_0,M_1]+[\wt L_1,M_0]=
dM_0-\p_T \wt L_0-t\sum_{i=1}^{2g}(\p_TU_i)\p_{\phi_i}\wt L_0,
\eeq
where $\phi_i$ are coordinates on fibers of the Jacobian bundle.
Here and below we skip for brevity index $a$, i.e. $t=t_a$, and $T=T_a$.
Let $\psi$ and $\psi^*$ be solutions of the adjoint systems of equations
\begin{eqnarray}
{}&\wt L_0\psi =\wt k\psi ,\ \ {}&\ \p_t\psi =M_0\psi, \label{ff1}\\
{}&\,\psi^*\wt L_0 =\wt k\psi^*,\ \  {}&\p_t\psi^* =-\psi^*M_0.\label{ff2}
\end{eqnarray}
Here $\psi^*$ is a vector-row, normalized by the condition $\psi^*\psi=1$.
>From (\ref{i1}, \ref{ff1}, \ref{ff2}) it follows that
\beq\label{i5}
\p_t\left(\psi^*L_1\psi\right)=-
\psi^*\left(\p_T \wt L_0-dM_0-t\sum_{i=1}^{2g}(\p_TU_i)\p_{\phi_i}\wt
L_0\right)\psi.
\eeq
>From the equations
\beq
\psi^{*}(\delta \wt L_0-\delta \wt k)\psi=-\psi^*(\wt L_0-\wt k)\delta \psi=0,
\eeq
and the normalization $\psi^*\psi=1$ it follows that
\beq\label{i2}
\psi^*\delta \wt L_0\psi=\delta \wt k.
\eeq
Variations of $\wt L_0$ with respect to the variables $\phi_i$ preserve the
spectral curve, i.e. for such variations $\delta \wt k=0$. Hence,
\beq
\psi^*\left(\p_{\phi_i}\wt L_0\right)\psi=0.
\eeq
Equation (\ref{i2}) also implies
\beq
\p_T \wt k=\psi^*(\p_T \wt L)\psi.
\eeq
Equations (\ref{ff1},\ref{ff2}) imply also
\beq
\psi^*(dM_0)\psi=-\psi^*M_0d\psi+\psi^*(\p_td\psi)=\p_t(\psi^*d\psi).
\eeq
Hence, (\ref{i5}) can be written as
\beq\label{i7}
\p_t(\psi^*L_1\psi)+\p_T \wt k-\p_t(\psi^*d\psi)=0.
\eeq
In \cite{kr0} it was shown that the solution $\psi$ of
equations (\ref{ff1}) is the conventional Baker-Akhiezer function
on $\wh \G$. Therefore, it can be written  explicitly in terms of the Riemann
theta-functions of the spectral curve.
In \cite{kr1} the original formulae were adapted for the averaging
procedure. In order to complete a proof of the Theorem, we do not need
these formulae in full. Let us present necessary facts.

The function $\psi=\psi(x,y;P)$, and the dual Baker-Akhiezer
function $\psi^*$ considered as functions of the real variables
$x,y$ can be represented in the form
\begin{eqnarray}
\psi(x,y;P)&=&\Phi(xU^r+yU^i+\zeta;P)
\exp\left(-\int^{P}xdv^r+ydv^i\right),
\label{ff3}\\
 \psi^*(x,y;P)&=&\Phi^*(xU^i+yU^r+\zeta;P)\exp\left(\ \int^P xdv^r+ydv^i
\right),
\label{ff4}
\end{eqnarray}
where $U^r,U^i$ are real $2g$-dimensional vectors, and for each
$P\in \wh \G$ the functions $\Phi(\zeta;P)$ and
$\Phi^*(\zeta;P)$ as functions of $2g$ real variables
$\zeta=(\zeta_1,\ldots,\zeta_{2g})$ have the following monodromy properties
\beq\label{ff5}
\Phi(\zeta+e_i;P)=w_i\Phi(\zeta),
\Phi^*(\zeta+e_i;P)=w_i^{-1}\Phi^{*}(\zeta;P),\ \ |w_i|=1.
\eeq
where $e_i$ are the basis vectors of $R^{2g}$.

The functions $\wt L_0$, $\psi$ and $\psi^*$
as functions of the complex variable $t=x+iy$ are meromorphic functions.
Therefore, if $L_1$ is uniformly bounded outside of some neighborhood of the
singularity locus, then the average value $<\p_t(\psi L_1\psi)>$ in $t$ of
the first term  in (\ref{i7}) equals zero
\beq
<f(t)>=\lim_{\Lambda_i\to \infty}\Lambda_i^{-1}\int_{0}^{\Lambda_i} fdt
\eeq
It is necessary to make a few remarks to clarify averaging procedure.
First of all, we assume that $0$ and $\Lambda_i$ are not in the locus.
The integral is taken along the path in the complex plane of the variable
$t$  which do not intersect singularities.

As follows from (\ref{ff3}-\ref{ff5}) the average value of the last term
in (\ref{i7}) does exist but depends on the direction in $t$-plane. If we
consider $t$, as a real variable then this average equals $-dv^r$.
For $t=iy$ it equals $-dv^i$, and therefore, the Theorem is proved.

\end{document}